\documentclass[preprint,amsmath,
tightenlines,nofootinbib,eqsecnum,showpacs,showkeys]{revtex4}

\usepackage{graphicx}

\newcommand{\cJ}{\mathcal{J}}
\newcommand{\cN}{\mathcal{N}}
\newcommand{\cV}{\mathcal{V}}
\newcommand{\bH}{\mathbf{H}}
\newcommand{\bS}{\mathbf{S}}
\newcommand{\rV}{\text{V}}
\newcommand{\rP}{\text{P}}
\newcommand{\rNP}{\text{NP}}
\newcommand{\rIm}{\text{Im}}
\newcommand{\tR}{\tilde{R}}
\newcommand{\tepsilon}{\tilde{\epsilon}}
\newcommand{\tr}{\text{tr}\,}

\begin{document}

%\begin{frontmatter}%ELSART

\title{$\cN$-fold Supersymmetry\\
 in Quantum Mechanics\\
 - Analyses of Particular Models -}

%REVTEX4
\author{Masatoshi Sato}
\email{msato@issp.u-tokyo.ac.jp}
\affiliation{The Institute for Solid State Physics,\\
 The University of Tokyo, Kashiwanoha 5-1-5,
 Kashiwa-shi, Chiba 277-8581, Japan}
\author{Toshiaki Tanaka}
\email{ttanaka@het.phys.sci.osaka-u.ac.jp}
\affiliation{Faculty of Integrated Human Studies,\\
 Kyoto University, Kyoto 606-8501, Japan}
\altaffiliation{Present address: Department of Physics,
Graduate School of Science, Osaka University, Toyonaka,
Osaka 560-0043, Japan}

%ELSART
%\author[Tokyo]{Masatoshi Sato\thanksref{Sato}}
%\author[Kyoto]{Toshiaki Tanaka\thanksref{Tanaka}}
%\address[Tokyo]{The Institute for Solid State Physics,
% The University of Tokyo, Kashiwanoha 5-1-5,
% Kashiwa-shi, Chiba 277-8581, Japan}
%\address[Kyoto]{Faculty of Integrated Human Studies,
% Kyoto University, Kyoto 606-8501, Japan}
%\thanks[Sato]{msato@issp.u-tokyo.ac.jp}
%\thanks[Tanaka]{ttanaka@phys.h.kyoto-u.ac.jp}

\date{\today}

\begin{abstract}
We investigate particular models which can be $\cN$-fold
supersymmetric at specific values of a parameter in the Hamiltonians.
The models to be investigated are a periodic potential and
a parity-symmetric sextic triple-well potential.
Through the quantitative analyses on the non-perturbative
contributions to the spectra by the use of the valley method, we show
how the characteristic features of $\cN$-fold supersymmetry which
have been previously reported by the authors can be observed.
We also clarify the difference between quasi-exactly solvable and
quasi-perturbatively solvable case in view of the dynamical property,
that is, dynamical $\cN$-fold supersymmetry breaking.
\end{abstract}

%REVTEX4
%REVTEX4

%\begin{keyword}%ELSART
%\end{keyword}%ELSART

\preprint{KUCP-0196}%REVTEX4

\maketitle%neELSART

%\end{frontmatter}%ELSART

\section{Introduction}
\label{sec:intro}

In our previous paper~\cite{AST2}, we have formulated in formal
and abstract ways \textit{$\cN$-fold supersymmetry\/}
in quantum mechanics~\cite{AKOSW2,ASTY,AST1,ANST1,ANST2}
and investigated general properties of the models which possess
this symmetry. $\cN$-fold supersymmetry is characterized by the
supercharges which are $\cN$-th order polynomials of momentum
and similar generalizations of supercharges were also investigated
in different contexts~\cite{And1,And2,And3,And4,Sam1,Sam2,Sam3,Sam4,%
Ply1,Ply2,Ply3,Ply4,Ros1,Kha1,Fer1,Fer2,DDT}.
We have shown that $\cN$-fold supersymmetric models have a lot of
significant properties similar to the ordinary supersymmetric ones%
~\cite{Wit1,Wit2,Sol1,Coo1} such as degenerate spectral structure between
bosonic states and fermionic ones, non-renormalization theorems for
the generalized Witten index and for a part of the spectra, and so on.
Furthermore, we have introduced the notion of
\textit{quasi-solvability\/} to identify an important aspect
of $\cN$-fold supersymmetry and have proved the equivalence
between $\cN$-fold supersymmetry and quasi-solvability.
Recently, we have further shown~\cite{ANST1} that \textit{Type A}
subclass of $\cN$-fold supersymmetry which was first introduced in
Ref.~\cite{AST1} is equivalent to the quasi-solvable models
constructed by $sl(2)$ generators~\cite{Turb}. Then, it has turned out
that the equivalence between them for special cases which had been
reported previously in Refs.~\cite{AKOSW2,Ply3,DDT} is generically holds.

Quasi-solvability means the existence of a finite dimensional
invariant subspace under the action of the Hamiltonian. As
a consequence, a part of the spectra can be solved by a finite
dimensional algebraic equation. In the case where the subspace is
physical, that is, $L^{2}$, these spectra give the \textit{true\/}
eigenvalues of the Hamiltonian. In this case, the system is often
called \textit{quasi-exactly solvable\/}~\cite{Turb,Shif,Ushv}. 
On the other hand, if the subspace is not physical, solvable spectra
only give \textit{perturbative\/} eigenvalues at most and thus we have
dubbed this case \textit{quasi-perturbatively solvable\/}~\cite{AST1}.
This distinction is quite important, especially in view of dynamical
$\cN$-fold supersymmetry breaking; $\cN$-fold supersymmetry is broken
dynamically if a system is quasi-perturbatively solvable while it
is not broken if a system is quasi-exactly solvable.

In this article, we analyze particular models which can be $\cN$-fold
supersymmetric more quantitatively. Previously in Ref.~\cite{AKOSW2},
an asymmetric quartic double-well potential was investigated in detail
by the valley method~\cite{Rowe,BaYu,Silv,AK,AW,HS,AKHSW,AKHOSW,AKOSW1}.
It was shown that the system possesses $\cN$-fold supersymmetry
at specific values of a parameter in the Hamiltonian where the leading
Borel singularity of the perturbative corrections for the first
$\cN$-th energies disappear. This result is a consequence of
a general property of $\cN$-fold supersymmetry, namely, the
non-renormalization theorem. It was also shown that the non-perturbative
corrections for the first $\cN$-th energies do not vanish even when
the system becomes $\cN$-fold supersymmetric. This result consistently
reflects the fact that, in the case of an asymmetric quartic double-well
potential, the solvable subspace is not physical, that is, the system
is quasi-perturbatively solvable; non-vanishing non-perturbative effects
break $\cN$-fold supersymmetry dynamically.

These observations show that combining with the formal discussions
in Ref.~\cite{AST2} quantitative analyses may give deeper and complement
understanding on dynamical properties of the $\cN$-fold supersymmetric
models. As in the case of the ordinary supersymmetric models,
non-perturbative analyses are quite important in the $\cN$-fold
supersymmetric case; dynamical $\cN$-fold supersymmetry
breaking can take place via purely non-perturbative effects, e.g.,
quantum tunneling. However, non-perturbative analyses are in general
quite non-trivial even in the simple one-dimensional quantum mechanics.
The valley method is one of the most successful tools for
this kind of purpose. So we fully employ it in this work.

The article organizes as follows. In the next section, we summarize
the general results and properties of the Type A subclass of
$\cN$-fold supersymmetry~\cite{AST2,AST1,ANST1,ANST2}.
In section \ref{sec:speci}, we develop particular cases of
the Type A models which are especially relevant for
the analyses in this article.
Sections \ref{sec:aexpo} and \ref{sec:acubi} are devoted to valley method
analyses on a periodic and a sextic triple-well potential, respectively.
We choose potentials to be investigated so that the systems can be
Type A $\cN$-fold supersymmetric at specific values of a parameter
involved in the potentials. The way of the choice enables us to clarify
the characteristic features of Type A $\cN$-fold supersymmetry.
The periodic potential is always quasi-exactly solvable when Type A
$N$-fold supersymmetric while the triple-well potential can be either
quasi-exactly or quasi-perturbatively solvable.
In both the cases we show that the disappearance of the leading Borel
singularity occurs. However, the non-perturbative corrections vanish
when and only when the systems are quasi-exactly solvable.
Finally, we give summary in the last section.

\section{General Properties of Type A $\cN$-fold Supersymmetry}
\label{sec:typeA}
First of all, we summarize the general results and properties of
Type A $\cN$-fold supersymmetry. For details, e.g., derivation of
the results, see Refs.~\cite{AST2,ANST1}. To define $\cN$-fold
supersymmetry, we introduce the following Hamiltonian $\bH_{\cN}$
and the $\cN$-fold supercharges,
\begin{eqnarray}
\bH_{\cN}=H_{\cN}^{-}(p, q)\psi\psi^{\dagger}
 +H_{\cN}^{+}(p, q)\psi^{\dagger}\psi ,
\label{eqn:defhm}
\end{eqnarray}
\begin{eqnarray}
Q_{\cN}=P_{\cN}^{\dagger}(p, q)\psi,\quad
 Q_{\cN}^{\dagger}=P_{\cN}(p, q)\psi^{\dagger}.
\label{eqn:defsc}
\end{eqnarray}
Here $\psi$ and $\psi^{\dagger}$ are fermionic coordinates
which satisfy,
\begin{eqnarray}
\{\psi , \psi\}=\{\psi^{\dagger}, \psi^{\dagger}\}=0,\quad
 \{\psi , \psi^{\dagger}\}=1,
\label{eqn:fmcom}
\end{eqnarray}
and are usually represented as the following $2\times 2$ matrix form:
\begin{eqnarray}
\psi =\left(
 \begin{array}{cc}
 0 & 0\\
 1 & 0
 \end{array}
\right),\quad
\psi^{\dagger}=\left(
 \begin{array}{cc}
 0 & 1\\
 0 & 0
 \end{array}
\right).
\label{eqn:mrepf}
\end{eqnarray}
The component of the $\cN$-fold supercharges $P_{\cN}$ is given
by an $\cN$-th order polynomial of $p=-id/dq$ and thus expressed as,
\begin{eqnarray}
P_{\cN}=p^{\cN}+w_{\cN -1}(q)p^{\cN -1}+\dots
 +w_{1}(q)p+w_{0}(q),
\label{eqn:gfnsc}
\end{eqnarray}
without any loss of generality. Then, the system is said to be
$\cN$-fold supersymmetric if the Hamiltonian $\bH_{\cN}$ commutes
with the $\cN$-fold supercharges $Q_{\cN}$ and $Q_{\cN}^{\dagger}$:
\begin{eqnarray}
[Q_{\cN}, \bH_{\cN}]=[Q_{\cN}^{\dagger}, \bH_{\cN}]=0.
\label{eqn:defns}
\end{eqnarray}

The Type A $\cN$-fold supersymmetry is characterized by a particular
class of the $\cN$-fold supercharges which can be expressed as
the following form:\footnote{Note that $W(q)$ in this paper is
the same as $\widetilde{W}(q)$ and \textit{not} as $W(q)$ in the
old notation of our previous paper~\cite{AST1,AST2,ANST1,ANST2}.
Since it is a bit troublesome to keep the unnecessary tilde, we omit it.}
\begin{eqnarray}
P_{\cN}&=&\left(D+i\frac{\cN -1}{2}E(q)\right)\left(D+i
 \frac{\cN -3}{2}E(q)\right)\dots\left(D-i\frac{\cN -1}{2}
 E(q)\right)\nonumber\\
&\equiv&\prod_{k=-(\cN -1)/2}^{(\cN -1)/2}
 \Bigl( D+ikE(q)\Bigr),\quad D=p-iW(q).
\label{eqn:anfsco0}
\end{eqnarray}
If we restrict $H_{\cN}^{\pm}$ to be the following
Schr\"odinger type,
\begin{eqnarray}
H_{\cN}^{\pm}=\frac{1}{2}p^{2}+V_{\cN}^{\pm}(q),
\label{eqn:schhm}
\end{eqnarray}
we can show~\cite{ANST1} that necessary and sufficient
conditions of the Hamiltonian (\ref{eqn:defhm}) with
(\ref{eqn:schhm}) to be Type A $\cN$-fold supersymmetric,
that is, to satisfy the relation (\ref{eqn:defns}),
are as the following:
\begin{subequations}
\label{eqns:anfsco}
\begin{eqnarray}
&&V_{\cN}^{\pm}(q)=\frac{1}{2}W(q)^2 +\frac{\cN^{\, 2}-1}{24}
 \Bigl(E(q)^2-2E'(q)\Bigr)\pm\frac{\cN}{2}W'(q),
\label{eqn:anfsco1}\\
&&\left(\frac{d}{dq}-E(q)\right)\frac{d}{dq}
 \left(\frac{d}{dq}+E(q)\right)W(q)=0 \quad (\cN\geq 2),
\label{eqn:anfsco2}\\
&&\left(\frac{d}{dq}-2E(q)\right)\left(\frac{d}{dq}-E(q)\right)
 \frac{d}{dq}\left(\frac{d}{dq}+E(q)\right)E(q)=0 \quad (\cN\geq 3).
\label{eqn:anfsco3}
\end{eqnarray}
\end{subequations}

\subsection{The Solvable Subspaces}
\label{ssec::solsub}
Owing to the relation Eq.~(\ref{eqn:defns}), the $\cN$-dimensional
vector spaces defined by
\begin{eqnarray}
\cV_{\cN}^{-}=\ker P_{\cN},\quad\cV_{\cN}^{+}=\ker P_{\cN}^{\dagger}
\label{eqn:defvs}
\end{eqnarray}
are invariant under the action of $H_{\cN}^{-}$ and $H_{\cN}^{+}$,
respectively. We can therefore define the matrices $\bS^{\pm}$
as follows:
\begin{eqnarray}
H_{\cN}^{\pm}\phi_{n}^{\pm}=\sum_{m=1}^{\cN}\bS_{n,m}^{\pm}\phi_{m}^{\pm},
\label{eqn:defma}
\end{eqnarray}
where $\phi^{\pm}$ are bases of the $\cV_{\cN}^{\pm}$, respectively.
It can be proved~\cite{AST2} for general $\cN$-fold supersymmetry
that the \textit{mother Hamiltonian} $\mathcal{H}_{\cN}$
defined by the anti-commutator of the supercharges can be expressed as,
\begin{eqnarray}
\mathcal{H}_{\cN}\equiv\frac{1}{2}\{ Q_{\cN}^{\dagger}, Q_{\cN}\}
 =\frac{1}{2}\left(
\begin{array}{cc}
 \det \mathbf{M}_{\cN}^{+}(H_{\cN}^{+})+p^{+}P_{\cN}^{\dagger} & 0\\
 0 & \det \mathbf{M}_{\cN}^{-}(H_{\cN}^{-})+p^{-\dagger}P_{\cN}
\end{array}
\right),
\label{eqn:rlmhoh}
\end{eqnarray}
where,
\begin{eqnarray}
\mathbf{M}_{\cN}^{\pm}(\lambda)=2(\lambda\mathbf{I}-\bS^{\pm}),
\label{eqn:defbM}
\end{eqnarray}
and $p^{\pm}$ are at most $(\cN -1)$-th order differential operators.
From the definition of the mother Hamiltonian (\ref{eqn:rlmhoh}),
the elements of the subspaces $\cV_{\cN}^{\pm}$ are also characterized
as the zero-modes of the mother Hamiltonian.

In the case of Type A, we can obtain analytic expressions for
these bases:
\begin{eqnarray}
\phi_{n}^{\pm}(q)=h(q)^{n-1}h'(q)^{-(\cN -1)/2}U(q)^{\pm 1},
\quad (n=1, \dots, \cN ),
\label{eqn:tabas}
\end{eqnarray}
where,\footnote{The definition of $U(q)$ is also
different from the one in our previous papers.}
\begin{eqnarray}
U(q)=e^{\int\! dq W(q)},
\label{eqn:defU}
\end{eqnarray}
and $h(q)$ is a solution of the following linear differential
equation:
\begin{eqnarray}
h''(q)-E(q)h'(q)=0,
\label{eqn:ODEh}
\end{eqnarray}
and thus generically given by,
\begin{eqnarray}
h(q)=c_{1}\int\! dq\, e^{\int\! dq\, E(q)}+c_{2}.
\label{eqn:gefoh}
\end{eqnarray}
The appearance of the two arbitrary constants $c_{1,2}$ in
Eq.~(\ref{eqn:gefoh}) reflects the fact that the spaces $\cV_{\cN}^{\pm}$
spanned by the bases Eq.~(\ref{eqn:tabas}) are invariant under any
linear transformations on $h(q)$.
With the aid of these bases Eq.~(\ref{eqn:tabas}), the components
of the matrices $\bS_{n,m}^{\pm}$ defined by Eq.~(\ref{eqn:defma}) can
be determined (for each fixed $n=1, \dots , \cN$) by the following
recurrence relations:
\begin{subequations}
\label{eqns:tpmt}
\begin{eqnarray}
\bS_{n,\,\cN -m}^{-}&=&\frac{P_{\cN -m-1}(H_{\cN}^{-}
 \phi_{n}^{-}-\sum_{k=\cN -m+1}^{\cN}\bS_{n, k}^{-}
 \phi_{k}^{-})}{P_{\cN -m-1}\phi_{\cN -m}^{-}},\\
\bS_{n,\,\cN -m}^{+}&=&\frac{P_{\cN -m-1}^{\dagger}(H_{\cN}^{+}
 \phi_{n}^{+}-\sum_{k=\cN -m+1}^{\cN}\bS_{n, k}^{+}
 \phi_{k}^{+})}{P_{\cN -m-1}^{\dagger}\phi_{\cN -n}^{+}},
\end{eqnarray}
\end{subequations}
for $m=1, \dots , \cN -1$ with the initial conditions,
\begin{eqnarray}
\bS_{n,\,\cN}^{-}=\frac{P_{\cN -1}H_{\cN}^{-}\phi_{n}^{-}}
 {P_{\cN -1}\phi_{\cN}^{-}},\quad
\bS_{n,\,\cN}^{+}=\frac{P_{\cN -1}^{\dagger}H_{\cN}^{+}
 \phi_{n}^{+}}{P_{\cN -1}^{\dagger}\phi_{\cN}^{+}}.
\label{eqn:inimt}
\end{eqnarray}

From Eq.~(\ref{eqn:defma}), the spectra $E_{n}^{\pm}$ of the
Hamiltonians $H_{\cN}^{\pm}$ in the subspaces
$\cV_{\cN}^{\pm}$ are given by,
\begin{eqnarray}
\det \mathbf{M}_{\cN}^{\pm}(E_{n}^{\pm})=0.
\label{eqn:detbM}
\end{eqnarray}
If $\phi_{n}(q)$'s are normalizable, linear combinations of
them which diagonalize the matrix $\bS$ are the \textit{true\/}
eigenstates of $H_{\cN}$. In this case, the system is often called
\textit{quasi-exactly solvable\/}~\cite{Turb,Shif,Ushv}. On the other
hand, if $\phi_{n}(q)$'s are not normalizable, they have, at most,
restricted meanings in the perturbation theory. In this case,
the spectra determined by Eq.~(\ref{eqn:detbM}) only give
perturbatively correct ones. For this reason,
we dub the case \textit{quasi-perturbatively
solvable\/}~\cite{AST2}. Then, $\cN$-fold supersymmetry of the total
system $\bH_{\cN}$ is dynamically broken when \textit{both\/} of
the systems $H_{\cN}^{\pm}$ are quasi-perturbatively solvable.
Otherwise, that is, at least one of the systems $H_{\cN}^{\pm}$
is quasi-exactly solvable, the elements of the corresponding solvable
subspace give the $\cN$-fold supersymmetric physical states and
therefore $\cN$-fold supersymmetry is preserved.

\subsection{A Non-renormalization Theorem}
\label{ssec:nonrth}
A kind of the non-renormalization theorem holds for the Type A
models. We first assume that we can set $W(0)=0$
by the redefinition of the origin of the coordinate $q$
and the energy. To define a perturbation
theory, we then introduce a coupling constant $g$ as,
\begin{eqnarray}
W(q)=\frac{1}{g}w(gq),\quad E(q)=g\, e(gq),
\label{eqn:defpt}
\end{eqnarray}
so that, in the leading order of $g$, the potential $V_{\cN}^{\pm}$
become harmonic with frequency $|w'(0)|$,
\begin{eqnarray}
V_{\cN}^{\pm}(q)=\frac{1}{2}w'(0)^{2}q^{2}+O(g).
\end{eqnarray}
From Eq.~(\ref{eqn:tabas}) with Eqs.~(\ref{eqn:defU}),
(\ref{eqn:gefoh}) and (\ref{eqn:defpt}), we can easily see that the
$\phi_{n}^{\pm}$ behave as,
\begin{eqnarray}
\phi_{n}^{\pm}(q)=U(0)^{\pm 1}\Bigl(q^{n-1}+O(g)\Bigr)
e^{\pm w'(0)q^{2}/2}.
\label{eqn:wfbh}
\end{eqnarray}
Here, we choose the two arbitrary constants $c_{1,2}$ in
Eq.~(\ref{eqn:gefoh}) as $h(0)=0$, $h'(0)=1$.
Thus, as far as $w'(0)>0\, (<0)$, all the $\phi_{n}^{-}
(\phi_{n}^{+})$ remain normalizable in \textit{any finite}
order of $g$ even if $\phi_{n}^{-} (\phi_{n}^{+})$ themselves
are \textit{not\/} normalizable. So, they stay the $\cN$-fold
supersymmetric vacua in any order of the perturbation theory
and therefore any perturbative corrections do not break $\cN$-fold
supersymmetry.

\section{Special Cases of Type A $\cN$-fold Supersymmetry}
\label{sec:speci}
In this section, we illustrate some special cases of the Type A
$\cN$-fold supersymmetry by using the general results obtained in
the previous section.

\subsection{Exponential Type Potentials}
\label{ssec:expon}
At first, we will consider the case where $E(q)=\lambda$
(non-zero constant). This is a trivial solution of
Eq.~(\ref{eqn:anfsco3}). From Eq.~(\ref{eqn:anfsco2}) we yield,
\begin{eqnarray}
W(q)=C_{1}e^{\lambda q}+C_{2}e^{-\lambda q}+C_{3}.
\label{eqn:Wexp}
\end{eqnarray}
In this case, the Hamiltonians and the supercharge are given by
\begin{eqnarray}
H_{\pm\cN}=\frac{1}{2}p^2 +\frac{1}{2}W(q)^2+\frac{\cN^{\, 2}-1}{24}
 \lambda^{2}\pm\frac{\cN}{2}W'(q),\quad
 P_{\cN}=\prod_{k=-(\cN -1)/2}^{(\cN -1)/2}\Bigl(D +ik\lambda\Bigr).
\label{eqn:Hmexp}
\end{eqnarray}
The function $h(q)$ can be chosen as,
\begin{eqnarray}
h(q)=\frac{e^{\lambda q}}{\lambda}.
\label{eqn:hexp}
\end{eqnarray}
Bases of the solvable subspaces $\cV_{\cN}^{\pm}$ are calculated as,
\begin{eqnarray}
\phi_{n}^{\pm}(q)=\frac{1}{\lambda^{n-1}}
 \exp\left[ -\frac{1}{2}(\cN -2n+1)\lambda q\pm C_{3}q
 \pm\frac{C_{1}}{\lambda}e^{\lambda q}\mp\frac{C_{2}}{\lambda}
 e^{-\lambda q}\right] .
\label{eqn:bsexp}
\end{eqnarray}
Thus, normalizability of $\phi_{n}^{\pm}$ depends on the values
of the constants $C_{i}$ and $\lambda$. For example, provided
that all the constants $C_{1}$, $C_{2}$ and $\lambda$
are non-zero real numbers, either $\phi^{+}$ or $\phi^{-}$ is
normalizable when $C_{1}C_{2}<0$ while both of $\phi^{\pm}$ are
not when $C_{1}C_{2}>0$. The correspondence between quasi-exact
solvability and $\cN$-fold supersymmetry in the case of the
exponential type potentials Eq.~(\ref{eqn:Wexp}) was recently
discussed in Ref.~\cite{Ply3}.

The non-zero matrix elements of $\bS^{\pm}$ can be calculated as
follows. The direct action of the
Hamiltonians Eq.~(\ref{eqn:schhm}), with the Type A potentials
Eq.~(\ref{eqn:anfsco1}), on the bases Eq.~(\ref{eqn:tabas}) reads,
\begin{eqnarray}
H_{\cN}^{\pm}\phi_{n}^{\pm}&=&-\frac{1}{2}(n-1)(n-2){h'}^{2}
 \phi_{n-2}^{\pm}+\frac{1}{2}(n-1)\Bigl[ (\cN -2)h''
 \mp 2W h'\Bigr]\phi_{n-1}^{\pm}\nonumber\\
&&-\frac{\cN -1}{12}\left[ (\cN -2)\left( E'+E^{2}
 \right)\mp 6\left( W'+E W\right)\right]\phi_{n}^{\pm}.
\label{eqn:haric1}
\end{eqnarray}
From Eqs.~(\ref{eqn:Wexp}) and (\ref{eqn:hexp}), the following
relations hold:
\begin{subequations}
\label{eqns:hplsexp}
\begin{eqnarray}
{h'}^{2}&=&\lambda^{2}h^{2},\\
h''&=&\lambda^{2}h,\\
E'+E^{2}&=&\lambda^{2},\\
Wh'&=&C_{1}\lambda^{2}h^{2}+C_{3}\lambda h+C_{2},\\
W'+EW&=&2C_{1}\lambda^{2}h+C_{3}\lambda.
\end{eqnarray}
\end{subequations}
Substituting the above relations (\ref{eqns:hplsexp})
for Eq.~(\ref{eqn:haric1}), we obtain,
\begin{subequations}
\label{eqns:Ssexp}
\begin{eqnarray}
&&\bS_{n,n-1}^{\pm}=\mp (n-1)C_{2},\\
&&\bS_{n,n}^{\pm}=-\frac{1}{12}\Bigl[ (\cN -1)(\cN -2)+6(n-1)(n-\cN )
 \Bigr]\lambda^{2}\pm\frac{1}{2}(\cN -2n+1)C_{3}\lambda,\\
&&\bS_{n,n+1}^{\pm}=\mp (n-\cN )C_{1}\lambda^{2}.
\end{eqnarray}
\end{subequations}
All the other matrix elements are zero.

The special choices $\lambda =ig$, $C_{1}=1/2ig$, $C_{2}=-1/2ig$
and $C_{3}=0$ lead to,
\begin{eqnarray}
W(q)=\frac{1}{g}\sin (gq),\quad E(q)=ig,
\label{eqn:expsp}
\end{eqnarray}
and correspond to the periodic potential in Ref.~\cite{ASTY}.
We note that Eq.~(\ref{eqn:expsp}) is incorporated with the
perturbation theory defined by Eq.~(\ref{eqn:defpt}).
We will later carry out non-perturbative analysis of this special
case in section \ref{sec:aexpo}.

\subsection{Sextic Oscillator Potentials}
\label{ssec:cubic}
Next, we will consider the case where,
\begin{eqnarray}
E(q)=\frac{1}{q-q_{0}}.
\label{eqn:Ecub}
\end{eqnarray}
This is also a solution of Eq.~(\ref{eqn:anfsco3}).
This special case corresponds to (one of)
the cubic type in Ref.~\cite{AST1}. The Hamiltonians and
the supercharge are given by,
\begin{eqnarray}
H_{\cN}^{\pm}&=&\frac{1}{2}p^2 +\frac{1}{2}W(q)^2
 +\frac{\cN^{\, 2}-1}{8(q-q_{0})^2}\pm\frac{\cN}{2}W'(q),
\label{eqn:Hmcub}\\
P_{\cN}&=&\prod_{k=-(\cN -1)/2}^{(\cN -1)/2}\left( D+i\frac{k}{q-q_{0}}
 \right),
\label{eqn:Pcub}
\end{eqnarray}
with,
\begin{eqnarray}
W(q)=C_{1}(q-q_{0})^3 +C_{2}(q-q_{0})+\frac{C_{3}}{q-q_{0}}.
\label{eqn:Wcub}
\end{eqnarray}
We note that the Hamiltonians (\ref{eqn:Hmcub}) are parity symmetric.
The function $h(q)$ can be chosen as,
\begin{eqnarray}
h(q)=\frac{(q-q_{0})^{2}}{2}.
\label{eqn:hcub}
\end{eqnarray}
Bases of the solvable subspaces $\cV_{\cN}^{\pm}$ are calculated as,
\begin{eqnarray}
\phi_{n}^{\pm}(q)=\frac{1}{2^{n-1}}
 (q-q_{0})^{2n-\frac{\cN}{2}-\frac{3}{2}\pm C_{3}}
 \exp\left[ \pm\frac{C_{1}}{4}(q-q_{0})^{4}\pm\frac{C_{2}}{2}
 (q-q_{0})^{2}\right].
\label{eqn:bscub}
\end{eqnarray}
Thus, either $\phi^{+}$ or $\phi^{-}$ is normalizable unless
$C_{1}=C_{2}=0$ and the corresponding system $H_{\cN}^{+}$ or
$H_{\cN}^{-}$ is quasi-exactly solvable. The correspondence
between quasi-exact solvability and $\cN$-fold supersymmetry
in the case of the sextic potential Eq.~(\ref{eqn:Hmcub}) was
recently pointed out in Ref.~\cite{DDT}.

The non-zero matrix elements of $\bS^{\pm}$ can be obtained
by Eq.~(\ref{eqn:haric1}).
From Eqs.~(\ref{eqn:Ecub}), (\ref{eqn:Wcub}) and (\ref{eqn:hcub}),
the following relations hold:
\begin{subequations}
\label{eqns:hplscub}
\begin{eqnarray}
{h'}^{2}&=&2h,\\
h''&=&1,\\
E'+E^{2}&=&0,\\
Wh'&=&4C_{1}h^{2}+2C_{2}h+C_{3},\\
W'+EW&=&8C_{1}h+2C_{2}.
\end{eqnarray}
\end{subequations}
Substituting the above relations (\ref{eqns:hplscub})
for Eq.~(\ref{eqn:haric1}) we obtain,
\begin{subequations}
\label{eqns:Sscub}
\begin{eqnarray}
&&\bS_{n,n-1}^{\pm}=\frac{1}{2}(n-1)\Bigl[(\cN -2n+2)\mp 2C_{3}
 \Bigr],\\
&&\bS_{n,n}^{\pm}=\pm(\cN -2n+1)C_{2},\\
&&\bS_{n,n+1}^{\pm}=\mp 4(n-\cN )C_{1}.
\end{eqnarray}
\end{subequations}
All the other matrix elements are zero.

If we rewrite Eq.~(\ref{eqn:Wcub}) as,
\begin{eqnarray}
W(q)=w(q)+\frac{C_{3}}{q-q_{0}},\quad
 w(q)=C_{1}(q-q_{0})^3 +C_{2}(q-q_{0}),
\label{eqn:cubicwa}
\end{eqnarray}
the potential parts $V_{\cN}^{\pm}(q)$ of the Hamiltonians
(\ref{eqn:Hmcub}) are, in terms of $w(q)$,
\begin{eqnarray}
V_{\cN}^{\pm}(q)=\frac{1}{2}w(q)^{2}+\frac{(2C_{3}\mp\cN -1)
 (2C_{3}\mp\cN +1)}{8(q-q_{0})^{2}}\pm\left(\frac{\cN}{2}\pm
 \frac{C_{3}}{3}\right)w'(q)+\frac{2}{3}C_{2}C_{3}.
\label{eqn:cubvw}
\end{eqnarray}
We note that in the cases when $C_{3}=(\cN\pm 1)/2$ and
$-(\cN\pm 1)/2$, one of the potential-pair $V_{\cN}^{\pm}(q)$
becomes a genuine sixth order polynomial:
\begin{subequations}
\label{eqns:gcubvw}
\begin{eqnarray}
V_{\cN}^{+}(q)&=&\frac{1}{2}w(q)^{2}+\frac{4\cN\pm 1}{6}w'(q)\quad
 \left( C_{3}=\frac{\cN\pm 1}{2}\right),\\
V_{\cN}^{-}(q)&=&\frac{1}{2}w(q)^{2}-\frac{4\cN\pm 1}{6}w'(q)\quad
 \left( C_{3}=-\frac{\cN\pm 1}{2}\right),
\end{eqnarray}
\end{subequations}
where irrelevant constant terms are omitted. Conversely, a sextic
anharmonic oscillator or a triple-well potential (with parity symmetry)
can be one of the Type A $\cN$-fold supersymmetric pair whenever
the potential can be put in one of the form of Eq.~(\ref{eqns:gcubvw}).
When $C_{3}=(\cN\pm 1)/2$, the bases Eq.~(\ref{eqn:bscub}) for
$V_{\cN}^{+}(q)$ read,
\begin{eqnarray}
\phi_{n}^{+}(q)=\frac{1}{2^{n-1}}(q-q_{0})^{2n-\frac{3}{2}
 \pm\frac{1}{2}}\exp\left[ \frac{C_{1}}{4}(q-q_{0})^{4}
 +\frac{C_{2}}{2}(q-q_{0})^{2}\right]\left(C_{3}=
 \frac{\cN\pm 1}{2}\right).
\label{eqn:bscbp1}
\end{eqnarray}
It is worth noting that the solvable subspace $\cV_{\cN}^{+}$ consists
of the states with definite parity (odd for $C_{3}=(\cN +1)/2$ and
even for $C_{3}=(\cN -1)/2$). We will later see close relation
between this fact and pattern of the non-perturbative spectral shifts.
When $C_{3}=-(\cN\pm 1)/2$, the bases Eq.~(\ref{eqn:bscub}) for
$V_{\cN}^{-}(q)$ are similarly,
\begin{eqnarray}
\phi_{n}^{-}(q)=\frac{1}{2^{n-1}}(q-q_{0})^{2n-\frac{3}{2}
 \pm\frac{1}{2}}\exp\left[ -\frac{C_{1}}{4}(q-q_{0})^{4}
 -\frac{C_{2}}{2}(q-q_{0})^{2}\right]\left( C_{3}=
 -\frac{\cN\pm 1}{2}\right).
\label{eqn:bscbp2}
\end{eqnarray}
Again, the subspace $\cV_{\cN}^{-}$ contains only odd-parity states
for $C_{3}=-(\cN +1)/2$ and only even-parity states for
$C_{3}=-(\cN -1)/2$.

\subsection{Quartic Oscillator Potentials}
\label{ssec:quadr}
In the next, we will consider the case where $E(q)=0$.
This is also a trivial solution of Eq.~(\ref{eqn:anfsco3}).
From Eq.~(\ref{eqn:anfsco2}) we yield,
\begin{eqnarray}
W(q)=C_{1}q^2 +C_{2}q +C_{3}.
\label{eqn:Wqua}
\end{eqnarray}
In this case, the Hamiltonians and the supercharge are given by
\begin{eqnarray}
H_{\pm\cN}=\frac{1}{2}p^2 +\frac{1}{2}W(q)^2\pm\frac{\cN}{2}W'(q),\quad
 P_{\cN}=D^{\cN}.
\label{eqn:Hmqua}
\end{eqnarray}
The function $h(q)$ reads,
\begin{eqnarray}
h(q)=q.
\label{eqn:hqua}
\end{eqnarray}
Bases of the solvable subspaces $\cV_{\cN}^{\pm}$ are calculated as,
\begin{eqnarray}
\phi_{n}^{\pm}(q)=q^{n-1}\exp\left[ \pm\frac{C_{1}}{3}q^{3}\pm
 \frac{C_{2}}{2}q^{2}\pm C_{3}q\right].
\label{eqn:bsqua}
\end{eqnarray}
Thus, both of $\phi^{\pm}$ are not normalizable and therefore
the system is quasi-perturbatively solvable as far as $C_{1}$ is
a non-zero real number. The relation between quasi-perturbative
solvability and $\cN$-fold supersymmetry in a special case of
the models was pointed out in Ref.~\cite{AKOSW2}.

The non-zero matrix elements of $\bS^{\pm}$ can be obtained
by Eq.~(\ref{eqn:haric1}).
From Eqs.~(\ref{eqn:Wqua}) and (\ref{eqn:hqua}), the following
relations hold:
\begin{subequations}
\label{eqns:hplsqua}
\begin{eqnarray}
{h'}^{2}&=&1,\\
Wh'&=&C_{1}h^{2}+C_{2}h+C_{3},\\
W'+EW&=&2C_{1}h+C_{2}.
\end{eqnarray}
\end{subequations}
Substituting the above relations (\ref{eqns:hplsqua})
for Eq.~(\ref{eqn:haric1}) we obtain,
\begin{subequations}
\label{eqns:Ssqua}
\begin{eqnarray}
&&\bS_{n,n-2}^{\pm}=-\frac{1}{2}(n-1)(n-2),\\
&&\bS_{n,n-1}^{\pm}=\mp (n-1)C_{3},\\
&&\bS_{n,n}^{\pm}=\pm\frac{1}{2}(\cN -2n+1)C_{2},\\
&&\bS_{n,n+1}^{\pm}=\mp (n-\cN )C_{1}.
\end{eqnarray}
\end{subequations}
All the other matrix elements are zero.

\section{Analysis of a Periodic Potential}
\label{sec:aexpo}
\subsection{The Valley Method}
\label{ssec:valley}
Before proceeding to show the results of the analyses,
we briefly review the valley method%
~\cite{Rowe,BaYu,Silv,AK,AW,HS,AKHSW,AKHOSW,AKOSW1}
which is employed in this research.
For more details about the method, see Ref.~\cite{AKOSW2}.

The main problem in quantum theories concerns with the evaluation
of the Euclidean partition function:
\begin{eqnarray}
Z=\mathcal{J}\int\mathcal{D}q\, e^{-S[q]}.
\label{eqn:partf}
\end{eqnarray}
Since the evaluation cannot be done exactly in general, one must
find out a proper method which enables one to get a good estimation
of the quantity. The semi-classical approximation is known to be one
of the most established methods. Especially, the uses of instantons
have been succeeded in analyzing non-perturbative aspects of
various quantum systems which have degenerate vacua~\cite{Cole}.
However, validity of the approximation comes into question when
the fluctuations around the classical configuration contain
a negative mode. Let us consider an asymmetric double-well
potential as a typical example. For this potential, there is
a so-called bounce solution as the classical solution which has
a negative mode in the fluctuations. The negative mode contributes
non-zero imaginary part of the spectra in the approximation, showing
instability of the system. Since the spectra of the model must be
real, the instability in the approximation must be \textit{fake}.

The appearance of a negative mode indicates that the classical action
does not give the minimum but rather a saddle point in the functional
space. In this case, one may expect that the quantity (\ref{eqn:partf})
is dominated by the configurations along the negative mode, which
may intuitively constitute a \textit{valley} in the functional space.
The valley method is a natural realization of this consideration.

At first, we give a geometrical definition of the valley in the
functional space $q(\tau)$~\cite{AK}:
\begin{eqnarray}
\frac{\delta}{\delta q(\tau)}\left[\frac{1}{2}\int\! d\tau'
 \left(\frac{\delta S[q]}{\delta q(\tau')}\right)^{2}
 -\lambda S[q]\right]=0.
\label{eqn:valley1}
\end{eqnarray}
The above definition (\ref{eqn:valley1}) can be interpreted as follows;
for each fixed ``height'' $S[q]$, the valley is defined at the point
where the norm of the gradient vector becomes extremal.
Introducing an auxiliary field $F(q)$, we can make the valley
equation (\ref{eqn:valley1}) a more perspicuous form:
\begin{subequations}
\label{eqns:valley2}
\begin{eqnarray}
\frac{\delta S[q]}{\delta q(\tau)}&=&F(\tau),\\
\int\! d\tau'\, D(\tau ,\tau')F(\tau')&=&\lambda F(\tau).
\end{eqnarray}
\end{subequations}
where the operator $D$ is defined as,
\begin{eqnarray}
D(\tau ,\tau')=\frac{\delta^{2} S[q]}{\delta q(\tau)\delta q(\tau')}.
\label{eqn:dfDop}
\end{eqnarray}
It is now evident that any solution of the equation of motion is
also a solution of the valley equation (\ref{eqns:valley2}) with
$F(\tau)\equiv 0$.

Next, we separate the integration along the valley line from the
whole functional integration. We parametrize the valley line by
a parameter $R$ and denote the valley configuration by $q_{R}(\tau)$.
We then define Faddeev-Popov determinant $\Delta[\varphi_{R}]$ by
the following:
\begin{eqnarray}
\int\! dR\,\delta\left(\int\! d\tau\,\varphi_{R}(\tau)G_{R}(\tau)
 \right)\Delta [\varphi_{R}]=1,
\label{eqn:FPdet}
\end{eqnarray}
where $\varphi_{R}(\tau)=q(\tau)-q_{R}(\tau)$ is the fluctuation
over which we will be doing Gaussian integrations, and $G_{R}(\tau)$
is the normalized gradient vector,
\begin{eqnarray}
G_{R}(\tau)=\frac{\delta S[q_{R}]}{\delta q_{R}(\tau)}\Bigg/
 \sqrt{\int\! d\tau'\,\left(\frac{\delta S[q_{R}]}{\delta q_{R}(\tau')}
 \right)^{2}}.
\label{eqn:gravc}
\end{eqnarray}
Inserting Eq.~(\ref{eqn:FPdet}) into the functional integral
(\ref{eqn:partf}), expanding the action $S[q]$ around
$\varphi_{R}(\tau)=0$ and integrating up to the second order term
in $\varphi_{R}(\tau)$, we finally obtain the one-loop order result:
\begin{eqnarray}
Z&=&\mathcal{J}\int\! dR\int\!\mathcal{D}q\,\delta\left(\int\! d\tau\,
 \varphi_{R}(\tau)G_{R}(\tau)\right)\Delta [\varphi_{R}] e^{-S[q]}
 \nonumber\\
&\simeq&\mathcal{J}\int\frac{dR}{\sqrt{2\pi\det' D_{R}}}
 \Delta [\varphi_{R}]e^{-S[q_{R}]},
\label{eqn:ptfiv}
\end{eqnarray}
where the Jacobian $\Delta [\varphi_{R}]$ is given by,
in this approximation,
\begin{eqnarray}
\Delta [\varphi_{R}]=\frac{d S[q_{R}]}{d R}\Bigg/
 \sqrt{\int\! d\tau'\,\left(\frac{\delta S[q_{R}]}{\delta q_{R}(\tau')}
 \right)^{2}}.
\label{eqn:FPapp}
\end{eqnarray}
In the above, $\det'$ denotes the determinant in the functional subspace
which is perpendicular to the gradient vector $G_{R}(\tau)$.
The valley equation (\ref{eqns:valley2}) ensure that the subspace
does not contain the eigenvector of the eigenvalue $\lambda$. Therefore,
we can safely perform the Gaussian integrations even when we encounter
a non-positive mode. The extension to the multi-dimensional valley,
which will be needed when there are multiple non-positive eigenvalues,
is straightforward.

In this article, we only deal with one-dimensional quantum mechanics
where the Euclidean action is given by,
\begin{eqnarray}
S[q]=\int\! d\tau\, \left[\frac{1}{2}\left(\frac{dq}{d\tau}
 \right)^{2}+V(q)\right].
\label{eqn:euact}
\end{eqnarray}
In this case, the valley equations (\ref{eqns:valley2}) are explicitly
written as,
\begin{subequations}
\label{eqns:valley3}
\begin{eqnarray}
-\frac{d^{2}q(\tau)}{d\tau^{2}}+V'(q)&=&F(\tau),\\
\left[ -\frac{d^{2}}{d\tau^{2}}+V''(q)\right]F(\tau)&=&\lambda F(\tau).
\end{eqnarray}
\end{subequations}

\subsection{Valley-Instantons}
\label{ssec:valins}
\begin{figure}[ht]
\begin{center}
\includegraphics[width=.6\textwidth]{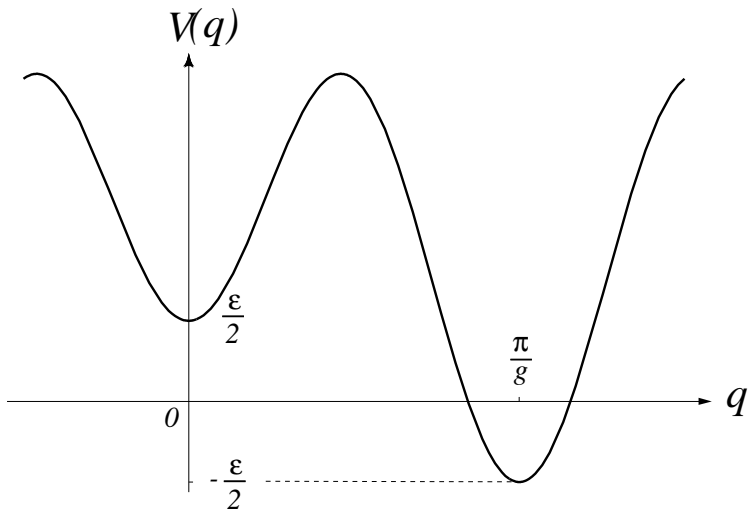}
\caption{The form of the periodic potential investigated
in this section.}
\label{fig:perid}
\end{center}
\end{figure}

At first, we will analyze a periodic potential.
The form of the potential to be analyzed is the following:
\begin{eqnarray}
V(q;\epsilon )=\frac{1}{2g^{2}}\sin^{2}(gq)+\frac{\epsilon}{2}\cos (gq).
\label{eqn:exppo}
\end{eqnarray}
This is a periodic potential with periodicity $2\pi /g$
(unless $\epsilon =0$) and has two local minima at $q=2k\pi /g$
and $q=(2k+1)/g$ $(k=0,\pm 1,\pm 2,\dots )$ in one period,
see Fig.~\ref{fig:perid}.

Comparing this potential with Eqs.~(\ref{eqn:Hmexp}) and
(\ref{eqn:expsp}), we find that the system has Type A $\cN$-fold
supersymmetry when
\begin{eqnarray}
\epsilon =\pm\cN .
\label{eqn:epvnfe}
\end{eqnarray}
Since the system is defined on a bounded region, all the bases
of the solvable space $\cV_{\cN}^{\pm}$ are normalizable. Thus, certain
linear combinations of them serve as physical eigenstates of the
Hamiltonian and $\cN$-fold supersymmetry is not broken dynamically.
Therefore, we may expect that the non-perturbative corrections
for certain $\cN$ physical states will vanish and the perturbation
series for the corresponding spectra will be convergent when
$\epsilon =\pm\cN$.

We note that the potential (\ref{eqn:exppo}) has the following
symmetry:
\begin{eqnarray}
V\left( q-\frac{\pi}{g};\epsilon\right)=V(q;-\epsilon ).
\label{eqn:exsym}
\end{eqnarray}
Therefore, we can restrict $\epsilon$ to be positive without
any loss of generality.

In the case of $\epsilon =0$, there are
(anti-)instanton solutions of the equation of motion which
describe the quantum tunneling between the neighboring vacua.
The instanton and anti-instanton which connect the two vacua
at $q=k\pi /g$ and $q=(k+1)\pi /g$ are given by,
\begin{subequations}
\label{eqns:expins}
\begin{eqnarray}
q_{0}^{(I)}(\tau -\tau_{0})&=&\frac{k\pi}{g}
 +\frac{1}{g}\arccos\Bigl( -\tanh (\tau -\tau_{0})\Bigr),\\
q_{0}^{(\bar{I})}(\tau -\tau_{0})&=&\frac{k\pi}{g}
 +\frac{1}{g}\arccos\Bigl( \tanh (\tau -\tau_{0})\Bigr).
\end{eqnarray}
\end{subequations}
When $\epsilon\ne 0$, the classical solutions drastically change into
the so-called bounce solutions which cause fake instability. On the
other hand, the solutions of the valley equation (\ref{eqns:valley3})
contain a continuously deformed (anti-)instanton which connects the
two non-degenerate local minima and is called
\textit{(anti-)valley-instanton}~\cite{AKOSW2}.

The solutions of the valley equation (\ref{eqns:valley3}) also
contain a family of the configurations, which tends to the trivial
vacuum configuration in the one limit and tends to well-separated
valley-instanton and anti-valley-instanton configuration in the
other limit. The latter configuration is called $I\bar{I}$-valley.
The bounce solution is also realized as an intermediate configuration
of this family, which is consistent with the fact that the solution
of the equation of motion is also a solution of the valley equations.
For details, see the numerical result in Ref.~\cite{AKOSW2}.
For the $I\bar{I}$-valley configuration,
it turns out that $|\lambda|\ll 1$ and thus the asymptotic form
of the configuration can be obtained by solving the valley equation
(\ref{eqns:valley3}) with perturbative expansion in $\lambda$:
\begin{eqnarray}
q(\tau)=q_{0}(\tau)+\lambda q_{1}(\tau)+\cdots ,
 \quad F(\tau)=\lambda F_{1}(\tau)+\lambda^{2} F_{2}(\tau)+\cdots .
\label{eqn:ptqaF4}
\end{eqnarray}
Indeed, if we denote the distance between the valley-instanton and
the anti-valley-instanton as $R$, the lambda is order
$\lambda\sim O(e^{-R})$ quantity. 
The action of the $I\bar{I}$-valley with the boundary condition
$q(\pm T/2)=2k\pi /g$ $(T\gg 1)$ is finally obtained as,
\begin{eqnarray}
S^{(I\bar{I})}(\tR)=S^{(\bar{I}I)}(\tR)
 =2S_{0}^{(I)}-\frac{\epsilon}{2}\tR+\frac{\epsilon}{2}(T-\tR )
 -\frac{8}{g^{2}}e^{-\tR}+O(e^{-2\tR}),
\label{eqn:Sibiex1}
\end{eqnarray}
while the one with $q(\pm T/2)=(2k+1)\pi /g$ $(T\gg 1)$ is,
\begin{eqnarray}
S^{(I\bar{I})}(R)=S^{(\bar{I}I)}(R)
 =2S_{0}^{(I)}+\frac{\epsilon}{2}R -\frac{\epsilon}{2}(T-R)
 -\frac{8}{g^{2}}e^{-R}+O(e^{-2R}),
\label{eqn:Sibiex2}
\end{eqnarray}
where $S_{0}^{(I)}$ denotes the Euclidean action of one (anti-)instanton
Eq.~(\ref{eqns:expins}) and amounts to,
\begin{eqnarray}
S_{0}^{(I)}=\frac{2}{g^{2}}.
\label{eqn:S0ex}
\end{eqnarray}
In Eqs.~(\ref{eqn:Sibiex1}) and (\ref{eqn:Sibiex2}),
the fourth term can be interpreted as
the interaction term between the valley-instanton and the
anti-valley-instanton. Therefore, the minus sign indicates that
the interaction is attractive.

The other type of the solutions emerges in this case, which is
asymptotically composed of two successive valley-instantons or
two successive anti-valley-instantons. We call them $II$-valley
and $\bar{I}\bar{I}$-valley, respectively. These configurations
do not appear in the case of double-well potentials since they
connect every other vacuum. The Euclidean action
of them with large separation $R$ can be also calculated in
the same way as,
\begin{eqnarray}
S^{(II)}(\tR)=S^{(\bar{I}\bar{I})}(\tR)
 =2S_{0}^{(I)}-\frac{\epsilon}{2}\tR+\frac{\epsilon}{2}(T-\tR)
 +\frac{8}{g^{2}}e^{-\tR}+O(e^{-2\tR}),
\label{eqn:Siiex1}
\end{eqnarray}
for the configuration with $q(-T/2)=2k\pi /g$ and
$q(T/2)=(2k\pm 2)\pi /g$ $(T\gg 1)$, and,
\begin{eqnarray}
S^{(II)}(R)=S^{(\bar{I}\bar{I})}(R)
 =2S_{0}^{(I)}+\frac{\epsilon}{2}R -\frac{\epsilon}{2}(T-R)
 +\frac{8}{g^{2}}e^{-R}+O(e^{-2R}),
\label{eqn:Siiex2}
\end{eqnarray}
for the one with $q(-T/2)=(2k+1)\pi /g$ and
$q(T/2)=(2k+1\pm 2)\pi /g$ $(T\gg 1)$.
Note that the sign of the fourth term is plus and thus the interaction
between the (anti-)valley-instantons in this case is repulsive.

\subsection{Analysis of Two-valley Sector}
\label{ssec:bogomo}
\begin{figure}[ht]
\begin{center}
\includegraphics[width=.6\textwidth]{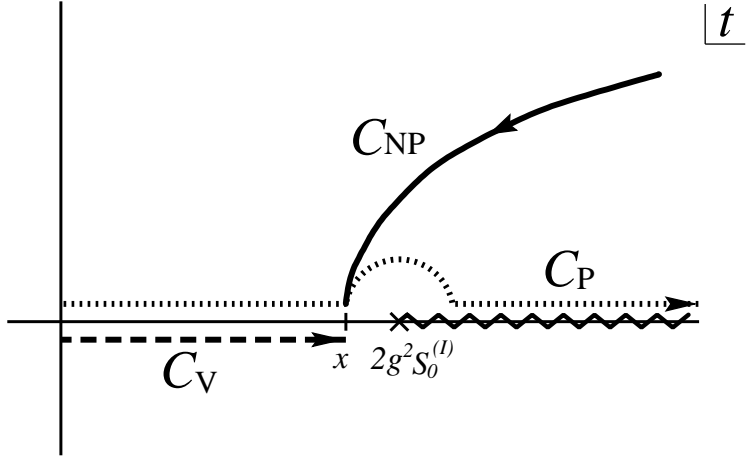}
\caption{Deformation of the contour $C_{\rV}$ to the sum of
$C_{\rP}$ and $C_{\rNP}$.}
\label{fig:defor}
\end{center}
\end{figure}

From the results on the $I\bar{I}$-valley, the contribution of
the $I\bar{I}$-valley to the partition function can be written
as the following form:
\begin{eqnarray}
Z^{(I\bar{I})}&=&\frac{4e^{-T/2}}{\pi g^{2}}\int_{0}^{T}\! dR\,
 (T-R)e^{-S^{I\bar{I}}(R)}\nonumber\\
&=&\frac{4e^{-T/2}}{\pi g^{2}}\int_{C_{\rV}}\! dt\,
 \mathcal{F}(t) e^{-t/g^{2}},
\label{eqn:Zibi1}
\end{eqnarray}
where we have changed the integration variable $R$ to
$t=g^{2}S^{I\bar{I}}(R)$ in the second line. The integration
contour $C_{\rV}$ is $[0,2g^{2}S_{0}^{(I)})$ and the integrand
$\mathcal{F}(t)$ has a singularity at $t=2g^{2}S_{0}^{(I)}$. The integral
Eq.~(\ref{eqn:Zibi1}) contains both the perturbative contribution
at $t\sim 0$ and the non-perturbative one at $t\sim 2g^{2}S_{0}^{(I)}$.
To separate the perturbative and the non-perturbative contribution,
we deform the contour $C_{\rV}$ to the sum of
$C_{\rP}$ and $C_{\rNP}$:
\begin{eqnarray}
Z^{(I\bar{I})}&=&\frac{4e^{-T/2}}{\pi g^{2}}\int_{C_{\rP}}\! dt\,
 \mathcal{F}(t) e^{-t/g^{2}}+\frac{4e^{-T/2}}{\pi g^{2}}
 \int_{C_{\rNP}}\! dt\, \mathcal{F}(t) e^{-t/g^{2}}\nonumber\\
&=&Z_{\rP}^{(I\bar{I})}(g^{2})
 +Z_{\rNP}^{(I\bar{I})}(g^{2}),
\label{eqn:dcpnp}
\end{eqnarray}
as is shown in Fig.~\ref{fig:defor}.

\begin{figure}[ht]
\begin{center}
\includegraphics[width=.6\textwidth]{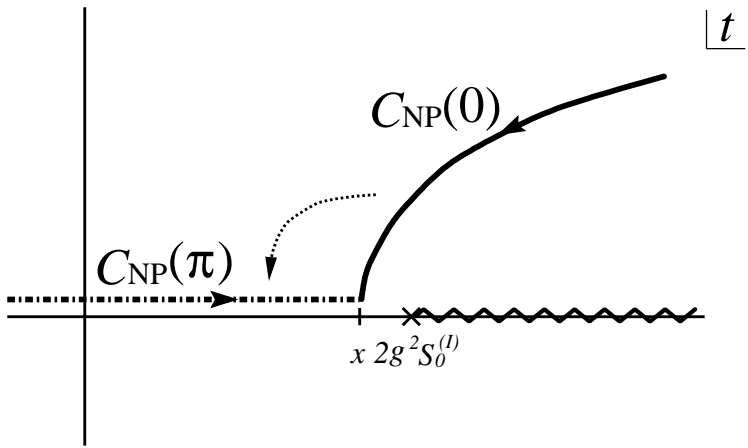}
\caption{The change of the contour $C_{\rNP}(\theta)$ as
$\theta$ is changed from zero to $\pi$.}
\label{fig:analc}
\end{center}
\end{figure}

We identify the first term
as the formal Borel summation of the perturbation series and the
second term as the non-perturbative contribution.
For the non-perturbative contribution, the following analytic
property holds.
If we perform the analytic continuation of
$Z_{\rNP}(|g^{2}|e^{i\theta})$ from $\theta =0$ to
$\theta =\pi$, the contour for $Z_{\rNP}$ changes from
$C_{\rNP}(0)$ to $C_{\rNP}(\pi)$, as shown in
Fig.~\ref{fig:analc}. In the weak coupling limit, the integral of
$C_{\rNP}(\pi)$ can be well-approximated by that of
$C_{\rV}$ because the dominant contribution of the integral
comes from $t\sim 2g^{2}S_{0}^{(I)}$. Therefore, in the case of
$g^{2}=|g^{2}|e^{i\pi}$ when the interaction between valley-instantons
is repulsive, the following relation holds approximately:
\begin{eqnarray}
Z_{\rNP}(|g^{2}|e^{i\pi})\simeq Z(|g^{2}|e^{i\pi}).
\label{eqn:bogrl1}
\end{eqnarray}
This relation coincides with what Bogomolny suggested heuristically
as a method of evaluation of the instanton--anti-instanton
contribution~\cite{Bogo}.

An immediate consequence of our decomposition of the perturbative
and non-perturbative contribution is,
\begin{eqnarray}
\rIm Z_{\rP}+\rIm Z_{\rNP}=0,
\label{eqn:imrel}
\end{eqnarray}
since $Z=Z_{\rP}+Z_{\rNP}$ is real. From the relation
above, the dispersion relation becomes~\cite{AKOSW2},
\begin{eqnarray}
Z_{\rP}(g^{2})&=&\frac{1}{2\pi i}\oint_{C_{g^{2}}}\! dz\,
 \frac{Z_{\rP}(z)}{z-g^{2}}\nonumber\\
%&\simeq&\frac{1}{\pi}\int_{0}^{\infty}\! dz\,
% \frac{\rIm Z_{\rP}(z)}{z-g^{2}}\nonumber\\
&\simeq&-\frac{1}{\pi}\sum_{r=0}^{\infty}g^{2r}
 \int_{0}^{\infty}\! dz\,\frac{\rIm Z_{\rNP}(z)}{z^{r+1}},
\label{eqn:Zdisper}
\end{eqnarray}
where $C_{g^{2}}$ is the counter around $z=g^{2}$ and we have neglected
the contribution from the singularities far from the origin. The last
line of Eq.~(\ref{eqn:Zdisper}) gives the relation between the
coefficients of the perturbation series and the imaginary part of
the non-perturbative contribution. Rewriting the above relation
in terms of the energy spectra, we find for the perturbative part
of the spectra $E_{\rP}(g^{2})=\sum_{r=0}^{\infty}a^{(r)}g^{2r}$,
that the coefficients $a^{(r)}$ can be estimated as,
\begin{eqnarray}
a^{(r)}=-\frac{1}{\pi}\int_{0}^{\infty}\! dg^{2}
 \frac{\rIm E_{\rNP}(g^{2})}{g^{2r+2}}.
\label{eqn:disrel}
\end{eqnarray}
The situation in the case of $\bar{I}I$-valley is completely
the same as that in the $I\bar{I}$-valley.

\begin{figure}[ht]
\begin{center}
\includegraphics[width=.6\textwidth]{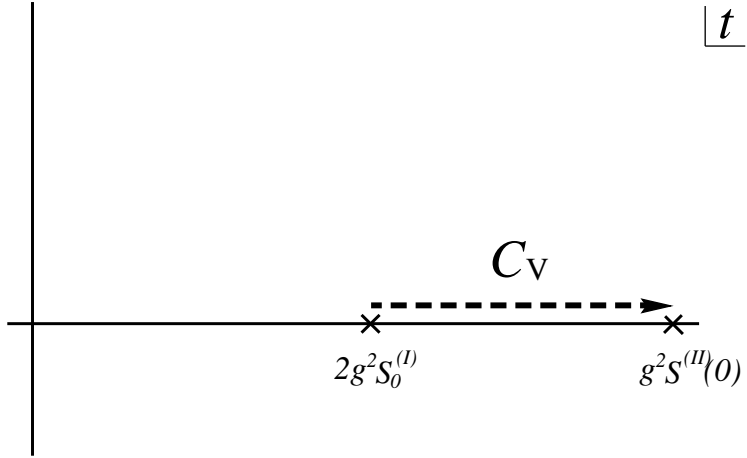}
\caption{The integration contour $C_{\rV}$ in the case
of the $II$-valley.}
\label{fig:icfII}
\end{center}
\end{figure}

On the other hand, the situation in the case of $II$-valley is
different, reflecting the fact that the $II$-valley configuration
cannot be deformed continuously to the trivial vacuum configuration.
The contribution of the $II$-valley to
the partition function has quite similar form to that of the
$I\bar{I}$-valley:
\begin{eqnarray}
Z^{(II)}=\frac{4e^{-T/2}}{\pi g^{2}}
 \int_{C_{\rV}}\! dt\,\mathcal{F}(t) e^{-t/g^{2}}.
\label{eqn:Zii1}
\end{eqnarray}
However, the integration contour is now
$C_{\rV}=(2g^{2}S_{0}^{(I)},g^{2}S^{(II)}(0)]$ and thus is
disconnected to the perturbative region $t\sim 0$, see
Fig.~\ref{fig:icfII}.

This means that Eq.~(\ref{eqn:Zii1}) contains only the
non-perturbative contribution,
\begin{eqnarray}
Z_{\rNP}(g^{2})=Z(g^{2}).
\label{eqn:bogrl2}
\end{eqnarray}
Therefore, we need not separate the integration as in the case of
the $I\bar{I}$-valley. As a consequence, the $II$-valley configuration
does not contribute to the imaginary part and also to the large
order behavior of the perturbation series, which will be confirmed
in the examples in sections \ref{sec:aexpo} and \ref{sec:acubi}.
Furthermore, since the interaction for the $II$-valley configuration
is repulsive, as has been observed in Eqs.~(\ref{eqn:Siiex1})
and (\ref{eqn:Siiex2}), the integral is dominated around
$t\sim 2g^{2}S_{0}^{(I)}$ and can be
approximated by the integral on the contour
$[2g^{2}S_{0}^{(I)},\infty )$ in the weak coupling limit.
Therefore, analytic continuation is not needed. The situation
in the case of the $\bar{I}\bar{I}$-valley is completely the same
as that in the $II$-valley.

\subsection{Multi-valley Calculus}
\label{ssec:mvcex}
\begin{figure}[ht]
\begin{center}
\includegraphics[width=.9\textwidth]{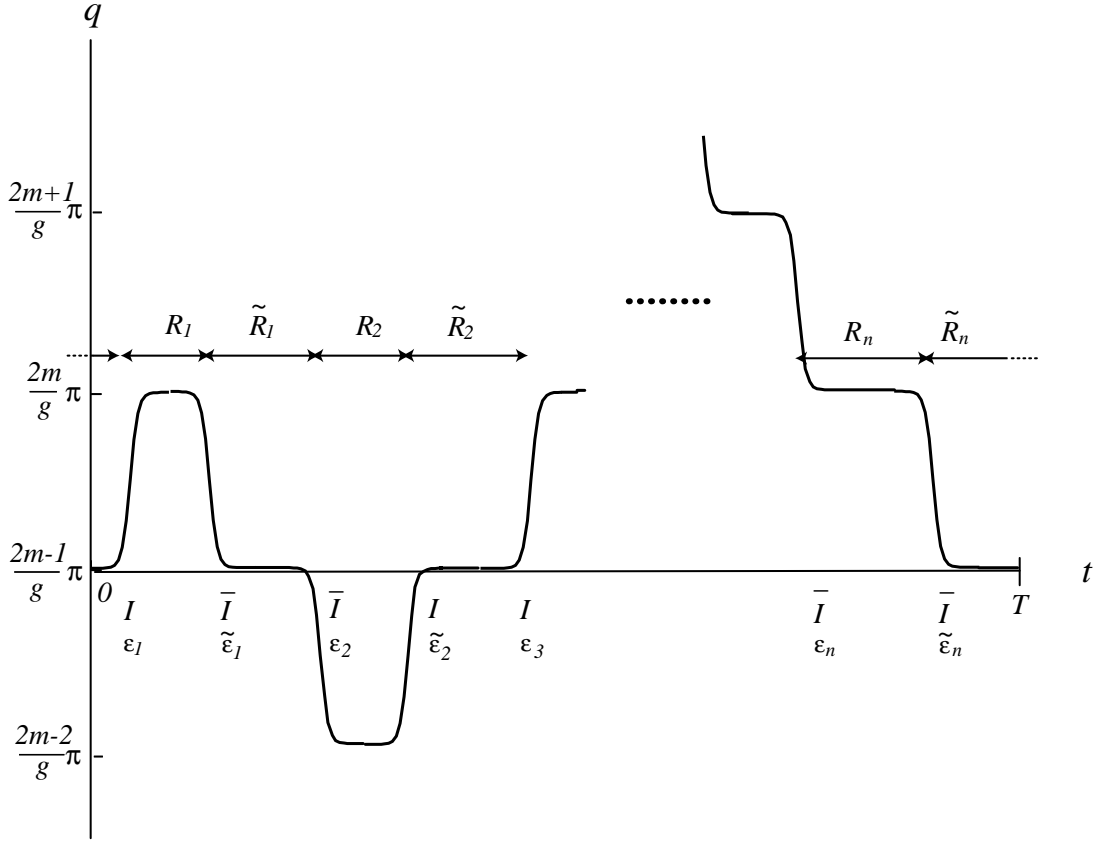}
\caption{The collective coordinates $R_{i}$ and $\tR_{i}$ for a
$2n$ valley-instantons configuration.}
\label{fig:mvcex}
\end{center}
\end{figure}

Utilizing the knowledge of the (anti-)valley-instantons and
the interactions between them obtained previously and
applying the manipulation discussed in subsection \ref{ssec:bogomo},
we will evaluate the partition function $Z=\tr e^{-HT}$ by summing
over those configurations made of several (anti-)valley-instantons
which satisfy a boundary condition in $T$.
The periodic boundary condition for a configuration $q(\tau )$
is in general given by
\begin{eqnarray}
q(\tau +T)=q(\tau ).
\label{eqn:perbc1}
\end{eqnarray}
For a system which has a periodic potential like Eq.~(\ref{eqn:exppo}),
however, the condition (\ref{eqn:perbc1}) can be relaxed and be
replaced with,
\begin{eqnarray}
q(\tau +T)=q(\tau )+\frac{2k\pi}{g}\quad (k=0, \pm 1, \pm 2, \dots ).
\label{eqn:perbc2}
\end{eqnarray}
This condition restricts the number of the valley-instantons to
be even, $2n$. The non-perturbative contributions from the
multi-valley configurations satisfying Eq.~(\ref{eqn:perbc2})
can be calculated by the extension of the technique developed
in Ref.~\cite{Zinn}. We divide the time interval $0\le\tau\le T$
into $n$ regions and put a valley-instanton pair on each of the
region. In order to distinguish what kind of pairs, we introduce
the indices $\epsilon_{i}$ and $\tepsilon_{i}$
for the $i$-th region$\pmod{n}$ as follows:
\renewcommand{\labelenumi}{(\roman{enumi})}
\begin{enumerate}
\item $\epsilon_{i}=1,\ \tepsilon_{i}=1\quad\text{for}
 \quad II\text{-valley}$,
\item $\epsilon_{i}=1,\ \tepsilon_{i}=-1\quad\text{for}
 \quad I\bar{I}\text{-valley}$,
\item $\epsilon_{i}=-1,\ \tepsilon_{i}=1\quad\text{for}
 \quad \bar{I}I\text{-valley}$,
\item $\epsilon_{i}=-1,\ \tepsilon_{i}=-1\quad\text{for}
 \quad \bar{I}\bar{I}\text{-valley}$.
\end{enumerate}
In this way, the allowed configurations for a given valley-instanton
number $2n$ are exhausted by the allowed combinations of the set
$\{\epsilon_{i},\tepsilon_{i}\}$ $(i=1, \dots , n)$.
Combining the results on well-separated valley-instanton pairs
with the above conventions, the well-separated multiple
valley-instanton action for given $n$ and
$\{\epsilon_{i},\tepsilon_{i}\}$ is expressed as,
\begin{eqnarray}
S_{n}&=&2nS_{0}^{(I)}+\frac{8}{g^{2}}\sum_{i=1}^{n}\epsilon_{i}
 \tepsilon_{i} e^{-R_{i}}+\frac{8}{g^{2}}\sum_{i=1}^{n}
 \tepsilon_{i}\epsilon_{i+1}e^{-\tR_{i}}\nonumber\\
&&+\frac{\epsilon}{2}
 \sum_{i=1}^{n} (R_{i}-{\tR}_{i})+i\frac{\vartheta}{2}\sum_{i=1}^{n}
 (\epsilon_{i}+\tepsilon_{i}),
\label{eqn:actex}
\end{eqnarray}
where $R_{i}$ is the distance between the $(2i-1)$-th and
$2i$-th (anti-)valley-instanton and $\tR_{i}$ the one between
the $2i$-th and the $(2i+1)$-th (anti-)valley-instanton$\pmod{n}$
and $\epsilon_{n+1}=\epsilon_{1}$, see Fig.~\ref{fig:mvcex}.

The sum of the contributions from the $2n$ valley-instantons
configuration can be written as,
\begin{eqnarray}
Z_{\rNP}=\sum_{n=1}^{\infty}\alpha^{2n}J_{n},
\label{eqn:Znpex}
\end{eqnarray}
where $\alpha^{2}$ denotes the contribution of the Jacobian and
the $R$-independent part of the determinant for one
valley-instanton-pair and is calculated as, in this case,
\begin{eqnarray}
\alpha^{2}=\frac{4}{\pi g^{2}}e^{-4/g^{2}}.
\label{eqn:defaex}
\end{eqnarray}
The term $J_{n}$ is given by,
\begin{eqnarray}
J_{n}&=&\frac{T}{n}\int_{0}^{\infty}\!\left(\prod_{i=1}^{n}dR_{i}\right)
 \left(\prod_{i=1}^{n}d\tR_{i}\right)\nonumber\\
&&\delta\left(\sum_{i=1}^{n}(R_{i}+\tR_{i})-T\right)
 \sum_{\epsilon_{i}, \tepsilon_{i}=\pm 1}\exp\left(
 -\frac{1+\epsilon}{2}\sum_{i=1}^{n}R_{i}-\frac{1-\epsilon}{2}
 \sum_{i=1}^{n}\tR_{i}\right.\nonumber\\
&&\left. -\frac{8}{g^{2}}\sum_{i=1}^{n}\epsilon_{i}
 \tepsilon_{i}e^{-R_{i}}-\frac{8}{g^{2}}\sum_{i=1}^{n}
 \tepsilon_{i}\epsilon_{i+1}e^{-\tR_{i}}-i\frac{\vartheta}{2}
 \sum_{i=1}^{n}(\epsilon_{i}+\tepsilon_{i})\right).
\label{eqn:dfJex}
\end{eqnarray}
To calculate the sum over the set $\{\epsilon_{i},\tepsilon_{i}\}$,
we introduce the following transfer matrices:
\begin{subequations}
\label{eqns:dftms}
\begin{eqnarray}
T(R_{i})&=&\left(
\begin{array}{cc}
\exp (-\frac{8}{g^{2}}e^{-R_{i}}-i\vartheta )
 &\exp (\frac{8}{g^{2}}e^{-R_{i}})\\
\exp (\frac{8}{g^{2}}e^{-R_{i}})
 &\exp (-\frac{8}{g^{2}}e^{-R_{i}}+i\vartheta )
\end{array}
\right),\\
\tilde{T}(\tR_{i})&=&\left(
\begin{array}{cc}
\exp (-\frac{8}{g^{2}}e^{-\tR_{i}})
 &\exp (\frac{8}{g^{2}}e^{-\tR_{i}})\\
\exp (\frac{8}{g^{2}}e^{-\tR_{i}})
 &\exp (-\frac{8}{g^{2}}e^{-\tR_{i}})
\end{array}
\right).
\end{eqnarray}
\end{subequations}
Then, using these matrices we have,
\begin{eqnarray}
J_{n}&=&\frac{T}{2\pi in}\int_{-i\infty -\eta}^{i\infty -\eta}\! ds\,
 e^{-Ts}\,\tr\left(\prod_{i=1}^{n}\int_{0}^{\infty}\! dR_{i}\,
 e^{(s-\frac{1}{2}-\frac{\epsilon}{2})R_{i}}T(R_{i})\int_{0}^{\infty}
 \! d\tR_{i}\, e^{(s-\frac{1}{2}+\frac{\epsilon}{2})\tR_{i}}
 \tilde{T} (\tR_{i})\right)\nonumber\\
&=&\frac{T}{2\pi in}\int_{-i\infty -\eta}^{i\infty -\eta}\! ds\,
 e^{-Ts}\,\tr\left[\mathcal{T} (s, \epsilon, \vartheta )^{n}\right],
\label{eqn:Jex1}
\end{eqnarray}
where,
\begin{eqnarray}
\mathcal{T} (s, \epsilon, \vartheta )=\left(
\begin{array}{cc}
K(s-\epsilon /2)e^{-i\vartheta}&I(s-\epsilon /2)\\
I(s-\epsilon /2)&K(s-\epsilon /2)e^{i\vartheta}
\end{array}
\right)\left(
\begin{array}{cc}
K(s+\epsilon /2)&I(s+\epsilon /2)\\
I(s+\epsilon /2)&K(s+\epsilon /2)
\end{array}
\right).
\label{eqn:defcT}
\end{eqnarray}
In the above, $K$ and $I$ are defined by,
\begin{subequations}
\label{eqns:dfKIex}
\begin{eqnarray}
K(s)&=&\int_{0}^{\infty}\! dR\, e^{(s-\frac{1}{2})R
 -\frac{8}{g^{2}}e^{-R}}
 \simeq\left(\frac{8}{g^{2}}\right)^{s-\frac{1}{2}}
 \Gamma\left( -s+\frac{1}{2}\right),\\
I(s)&=&\int_{0}^{\infty}\! dR\, e^{(s-\frac{1}{2})R
 +\frac{8}{g^{2}}e^{-R}}
 \simeq\left( -\frac{8}{g^{2}}\right)^{s-\frac{1}{2}}
 \Gamma\left( -s+\frac{1}{2}\right),
\end{eqnarray}
\end{subequations}
where the manipulation explained in subsection \ref{ssec:bogomo} is
utilized for estimating each of the integration. The calculation
of the trace in Eq.~(\ref{eqn:Jex1}) can be done by diagonalizing
$\mathcal{T}$. If we denote the two eigenvalues of $\mathcal{T}$
as $t_{\pm}$, we immediately yield,
\begin{eqnarray}
J_{n}=\frac{T}{2\pi in}\int_{-i\infty -\eta}^{i\infty -\eta}\! ds\,
 e^{-Ts}\Bigl( t_{+}(s)^{n}+t_{-}(s)^{n}\Bigr).
\label{eqn:Jex2}
\end{eqnarray}
From Eq.~(\ref{eqn:defcT}), $t_{\pm}$ is evaluated as,
\begin{eqnarray}
t_{\pm}(s)&=&K(s+\epsilon /2)K(s-\epsilon /2)\cos\vartheta
 +I(s+\epsilon /2)I(s-\epsilon /2)\nonumber\\
&&\pm\biggl\{\Bigl[ I(s+\epsilon /2)^{2}
 -K(s+\epsilon /2)^{2}\Bigr] K(s-\epsilon /2)^{2}\sin^{2}\vartheta
 \nonumber\\
&&+\Bigl[ K(s+\epsilon /2)I(s-\epsilon /2)+I(s+\epsilon /2)
 K(s-\epsilon /2)\cos\vartheta\Bigr]^{2}\biggr\}^{1/2}.
\label{eqn:dfpmt}
\end{eqnarray}
Finally, combining Eqs.~(\ref{eqn:Znpex}) and (\ref{eqn:Jex2})
we obtain the non-perturbative contribution to the partition function:
\begin{eqnarray}
Z_{\rNP}=-\frac{T}{2\pi i}\int_{-i\infty-\eta}^{i\infty-\eta}
 \! ds\, e^{-Ts}\ln\Bigl( 1-\alpha^{2}t_{+}(s)\Bigr)
 \Bigl( 1-\alpha^{2}t_{-}(s)\Bigr).
\label{eqn:Zex1}
\end{eqnarray}

\subsection{Non-perturbative Contributions}
\label{ssec:npcex}
\begin{figure}[ht]
\begin{center}
\includegraphics[width=.8\textwidth]{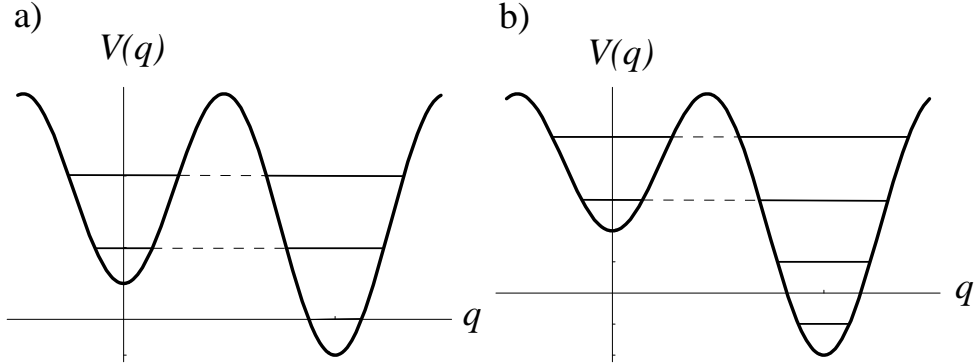}
\caption{Degeneracies of the harmonic spectra for (a)$\epsilon =1$
and (b)$\epsilon =2$.}
\label{fig:dgexp}
\end{center}
\end{figure}

From the results in Eqs.~(\ref{eqns:dfKIex}), (\ref{eqn:dfpmt}) and
(\ref{eqn:Zex1}), the non-perturbative contributions to
the spectra are determined by the following equation:
\begin{eqnarray}
\alpha^{2}\beta_{\pm}(E, \epsilon , \vartheta )\left(\frac{8}{g^{2}}
 \right)^{(E-\frac{1}{2})2}\Gamma\left(-E+\frac{1}{2}
 -\frac{\epsilon}{2}\right)\Gamma\left(-E+\frac{1}{2}
 +\frac{\epsilon}{2}\right)=1,
\label{eqn:spdex}
\end{eqnarray}
where,
\begin{eqnarray}
\beta_{\pm}(E, \epsilon , \vartheta )
 &=&\cos\vartheta +(-)^{(E-\frac{1}{2})2}\nonumber\\
&&\pm\sqrt{(-)^{(E-\frac{1}{2}-\frac{\epsilon}{2})2}
 +(-)^{(E-\frac{1}{2}+\frac{\epsilon}{2})2}
 +(-)^{(E-\frac{1}{2})2}2\cos\vartheta -\sin^{2}\vartheta}.
\label{eqn:defbex}
\end{eqnarray}
We will solve the above equation by the series expansion in $\alpha$:
\begin{eqnarray}
E_{n_{\pm}}=E_{n_{\pm}}^{(0)}
 +\alpha E_{n_{\pm}}^{(1)}+\alpha^{2} E_{n_{\pm}}^{(2)}+\cdots ,
 \quad E_{n_{\pm}}^{(0)}=n_{\pm}+\frac{1}{2}\pm\frac{\epsilon}{2} ,
\label{eqn:piaex}
\end{eqnarray}
where $E_{n_{+}}$ stands for the spectra corresponding to, in the
limit $g\to 0$, the eigenfunctions of the shallower potential wells
and $E_{n_{-}}$ for the ones corresponding to the eigenfunctions
of the deeper potential wells. For $\epsilon\ne\cN$
$(\cN =1, 2, 3, \dots )$,
the 1st order contributions vanish and the leading 2nd order
contributions are calculated as follows:
\begin{eqnarray}
E_{n_{\pm}}^{(2)}=2\left[\cos\vartheta +(-)^{\pm\epsilon}\right]
 \frac{(-)^{n_{\pm}+1}}{n_{\pm}!}\left(\frac{8}{g^{2}}
 \right)^{2n_{\pm}\pm\epsilon}\Gamma (-n_{\pm}\mp\epsilon ).
\label{eqn:e2ex1}
\end{eqnarray}
For $\epsilon=\cN$ $(\cN =1, 2, 3, \dots )$,
all the harmonic spectra $E_{n_{+}}^{(0)}$ of the shallower wells
and the higher harmonic spectra $E_{n_{-}}^{(0)}$ of the deeper wells
degenerate for $n_{-}=n_{+}+\cN$, see Fig.~\ref{fig:dgexp} for
$\epsilon=\cN =1,2$.
Between these degenerate states, resonant tunneling enhances
the non-perturbative corrections, and results in order $\alpha^{1}$
contributions to the spectra $E_{n_{-}}$ and $E_{n_{+}}$
with $n_{-}=n_{+}+\cN$:
\begin{eqnarray}
E_{n_{\pm}}^{(1)}&=&\mp\sqrt{\frac{2\left[ 1+(-)^{\cN}\cos\vartheta
 \right]}{n_{+}!\, n_{-}!}\left(\frac{8}{g^{2}}\right)^{n_{+}+n_{-}}},
\label{eqn:e1ex1}\\
E_{n_{\pm}}^{(2)}&=&\frac{E_{n_{\pm}}^{(1)\, 2}}{2}\left[
 \frac{2\ln (-)}{1+(-)^{\cN}\cos\vartheta}+2\ln \left(
 \frac{8}{g^{2}}\right)-\psi (n_{+}+1)-\psi (n_{-}+1)\right],
\label{eqn:e2ex2}
\end{eqnarray}
where $\psi(z)=d \ln\Gamma(z)/dz$ is the digamma function.
For the other spectra, say, the lower $E_{n_{-}}$ with
$n_{-}<\cN$, the contributions are the same as Eq.~(\ref{eqn:e2ex1}).
We see from Eq.~(\ref{eqn:e2ex1}) the non-perturbative corrections
for these lower $E_{n_{-}}$ vanish, at least up to order $\alpha^{2}$,
at $\vartheta =0\, (\pi)$ when $\epsilon=\cN$ is odd(even) and thus,
from Eq.~(\ref{eqn:epvnfe}), the system has $\cN$-fold supersymmetry.
This means that when the system is $\cN$-fold supersymmetric with
odd(even) $\cN$, among the physical
states for each of the lower $\cN$ spectral bands, the state which
satisfies the periodic(anti-periodic) boundary condition does not
receive non-perturbative correction. From Eq.~(\ref{eqn:bsexp}) in
this case, these physical states are surely the elements of the
solvable subspace $\cV^{-}$. Therefore, the results are consistent
with the fact that $\cN$-fold supersymmetry in this case is not
broken dynamically.

Finally, we make a remark on the resultant equation
(\ref{eqn:spdex}). The origin of the disappearance of the
non-perturbative corrections discussed above comes from the factor
$\beta_{\pm}$ in Eq.~(\ref{eqn:spdex}). Indeed, to make the l.h.s.\ of
Eq.~(\ref{eqn:spdex}) finite when $\beta_{\pm}=0$, the gamma
functions must diverge adequately. This happens only when
$E=E_{n_{\pm}}^{(0)}$ for certain values of $n_{\pm}$,
and therefore the non-perturbative corrections must vanish.\footnote{
Note that the argument does not depend on the perturbative
expansion in $\alpha$.} From the derivation of Eq.~(\ref{eqn:spdex}),
we can see that the appearance of the factor $\beta_{\pm}$ is
achieved by taking into account both of the repulsive and attractive
interactions between the valley-instantons properly. Therefore,
we guess that naive application of the dilute-gas approximation
can hardly lead to the correct results even for the ground-state
energy.

\subsection{Large Order Behavior of the Perturbation Series}
\label{ssec:lobex}
The large order behavior of the perturbation series in $g^{2}$
for the spectra can be estimated by the same way as in the case
of the double-well potential. From the non-perturbative
contributions Eqs.~(\ref{eqn:e2ex1})--(\ref{eqn:e2ex2}), we can
easily see that the imaginary parts of them are continuous, at
least up to the order $\alpha^{2}$, in $\epsilon$ and yield,
\begin{eqnarray}
\rIm E_{n_{\pm}}\sim -\alpha^{2}\frac{2\pi}
 {n_{\pm}!\, \Gamma (n_{\pm}+1\pm\epsilon )}\left(
 \frac{8}{g^{2}}\right)^{2n_{\pm}\pm\epsilon},
\label{eqn:imeex}
\end{eqnarray}
which are valid for arbitrary $\epsilon$. Then, if we expand
the spectra in power of $g^{2}$ such that,
\begin{eqnarray}
E_{n_{\pm}}=E_{n_{\pm}}^{(0)}+\sum_{r=1}^{\infty}
 a_{n_{\pm}}^{(r)}g^{2r},
\label{eqn:dfptex}
\end{eqnarray}
the large order behavior of the coefficients $a^{(r)}$ for
sufficiently large $r$ are calculated as, using Eqs.~(\ref{eqn:disrel})
and (\ref{eqn:imeex}),
\begin{eqnarray}
a_{n_{\pm}}^{(r)}\sim A_{n_{\pm}}(\epsilon)\, 4^{-r}\,\Gamma\left(
 r+2n_{\pm}+1\pm\epsilon\right),
\label{eqn:lobex}
\end{eqnarray}
where,
\begin{eqnarray}
A_{n_{\pm}}(\epsilon)=\frac{2}{\pi}\frac{2^{2n_{\pm}\pm\epsilon}}
 {n_{\pm}!\,\Gamma\left( n_{\pm}+1\pm\epsilon\right)}.
\label{eqn:lobpex}
\end{eqnarray}
Equation (\ref{eqn:lobex}) shows that the perturbative
coefficients diverge factorially unless the prefactor $A(\epsilon )$
vanish. From Eq.~(\ref{eqn:lobpex}), we can find the disappearance
of the leading divergence takes place only when $\epsilon =\cN$
$(\cN =1, 2, 3, \dots )$.
Comparing the results with Eq.~(\ref{eqn:epvnfe}) and taking
the symmetry (\ref{eqn:exsym}) into account, we see
that the above cases completely coincide with the case where
the system possesses Type A $\cN$-fold supersymmetry. Therefore,
the results of the valley method analyses are consistent with
a consequence of Type A $\cN$-fold supersymmetry, that is,
the non-renormalization theorem.

\section{Analysis of a Triple-well Potential}
\label{sec:acubi}
\begin{figure}[ht]
\begin{center}
\includegraphics[width=.6\textwidth]{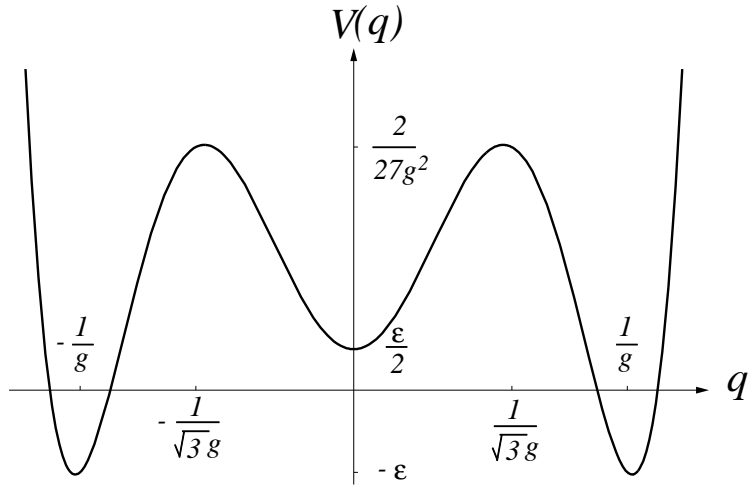}
\caption{The form of the triple-well potential investigated
in this section.}
\label{fig:tripl}
\end{center}
\end{figure}

In this section, we will analyze a sextic triple-well potential.
The form of the potential to be analyzed is the following:
\begin{eqnarray}
V(q)=\frac{1}{2}q^{2}(1-g^{2}q^{2})^{2}+\frac{\epsilon}{2}
 (1-3g^{2}q^{2}).
\label{eqn:cubpo}
\end{eqnarray}
This has three local minima at $q=0$ and $q\simeq\pm 1/g$ for
$\epsilon g^{2}\ll 1$, see Fig.~\ref{fig:tripl}.

Comparing this potential with Eq.~(\ref{eqns:gcubvw}), $C_{1}$,
$C_{2}$ and $q_{0}$ in $w(q)$ being,
\begin{eqnarray}
C_{1}=-g^{2},\quad C_{2}=1,\quad q_{0}=0,
\label{eqn:C1C2w}
\end{eqnarray}
we can easily see that the system has Type A $\cN$-fold
supersymmetry when
\begin{eqnarray}
\epsilon =\pm\frac{4\cN\pm 1}{3}.
\label{eqn:epvnfc}
\end{eqnarray}
More precisely, the system becomes one of the $\cN$-fold supersymmetric
pair $H_{\cN}^{\pm}$; $H_{\cN}^{+}$ when $\epsilon =(4\cN\pm 1)/3$
and $H_{\cN}^{-}$ when $\epsilon =-(4\cN\pm 1)/3$.
As has been explained in subsection \ref{ssec:cubic}, $\cN$-fold
supersymmetry does not break in the cubic type because either bases
of the solvable subspace $\cV_{\cN}^{+}$ or those of $\cV_{\cN}^{-}$
are normalizable in general.
Since $C_{1}<0$ in this case, bases of the solvable subspace
$\cV_{\cN}^{+}$ are normalizable and physical while those of
$\cV_{\cN}^{-}$ are not. Therefore, we may expect that
the non-perturbative corrections for certain $\cN$ states will
vanish when $\epsilon =(4\cN\pm 1)/3$ while those will not when
$\epsilon =-(4\cN\pm 1)/3$, though the perturbation series for
the corresponding spectra will be in both the cases convergent.

As far as we know, little has been investigated for triple-well
potentials. We have found only two references~\cite{SYLee,Casah}
on the subject, both of which employed the dilute-gas approximation.
However, from the consideration mentioned at the end of subsection
\ref{ssec:npcex}, we intend to analyze beyond the dilute-gas
approximation using the same technique of the valley method as
in section \ref{sec:aexpo}.\footnote{Indeed, the results of
Ref.~\cite{SYLee,Casah} for the lowest three energies are different
from ours and are not consistent with $\cN$-fold supersymmetry.}

In the case of $\epsilon =0$, the three local minima of
the potential have the same potential value. Thus, there are
(anti-)instanton solutions of the equation of motion which
describe the quantum tunneling between the neighboring vacua:
\begin{eqnarray}
q_{0}^{(I)}(\tau -\tau_{0})=\pm\frac{1}{g}\frac{1}
 {(1+e^{\mp 2(\tau -\tau_{0})})^{1/2}},\quad
q_{0}^{(\bar{I})}(\tau -\tau_{0})=\pm\frac{1}{g}\frac{1}
 {(1+e^{\pm 2(\tau -\tau_{0})})^{1/2}}.
\label{eqn:cubins}
\end{eqnarray}
When $\epsilon\ne 0$, the solutions of the valley equation now
become the (anti-)valley-instantons. In this case, there are three
kinds of the solutions of the valley equation which are asymptotically
composed of two (anti-)valley-instantons. Contrary to the periodic
potential in section \ref{sec:aexpo}, there are two different
$I\bar{I}$-valley or $\bar{I}I$-valley configurations in this case
since the curvature at the central potential bottom (at $q=0$) is
different, even at the leading order of $g^{2}$, from the one
at the side potential bottoms (at $q\simeq\pm 1/g$); the $I\bar{I}$
($\bar{I}I$)-valley which satisfy $q(\pm T/2)=0$ $(T\gg 1)$ are
different from the ones which satisfy $q(\pm T/2)\simeq 1/g$ or $-1/g$
$(T\gg 1)$.
The Euclidean action of the former with large separation $R$ can be
calculated by the perturbative expansion in
$\lambda\sim O(e^{-2R})$ as follows:
\begin{eqnarray}
S^{(I\bar{I})}(R)=S^{(\bar{I}I)}(R)
 =2S_{0}^{(I)}-\epsilon R+\frac{\epsilon}{2}(T-R)
 -\frac{1}{g^{2}}e^{-2R}+O(e^{-4R}),
\label{eqn:Sibi1c}
\end{eqnarray}
while the one of the latter with large separation $\tR$ can be calculated
in the same way as,
\begin{eqnarray}
S^{(I\bar{I})}(\tR)=S^{(\bar{I}I)}(\tR)
 =2S_{0}^{(I)}+\frac{\epsilon}{2}\tR -\epsilon (T-\tR )
 -\frac{2}{g^{2}}e^{-\tR}+O(e^{-2\tR}),
\label{eqn:Sibi2c}
\end{eqnarray}
where $S_{0}^{(I)}$ denotes the Euclidean action of one (anti-)instanton
Eq.~(\ref{eqn:cubins}) and amounts to,
\begin{eqnarray}
S_{0}^{(I)}=\frac{1}{4g^{2}}.
\label{eqn:S0cu}
\end{eqnarray}
The other type is $II$-valley or $\bar{I}\bar{I}$-valley.
The Euclidean action
of them with large separation $\tR$ can be also calculated in
the same way as,
\begin{eqnarray}
S^{(II)}(\tR)=S^{(\bar{I}\bar{I})}(\tR)
 =2S_{0}^{(I)}+\frac{\epsilon}{2}\tR-\epsilon (T-\tR )
 +\frac{2}{g^{2}}e^{-\tR}+O(e^{-2\tR}).
\label{eqn:Siicu}
\end{eqnarray}

\subsection{Multi-valley Calculus}
\label{ssec:mvccu}
\begin{figure}[ht]
\begin{center}
\includegraphics[width=.9\textwidth]{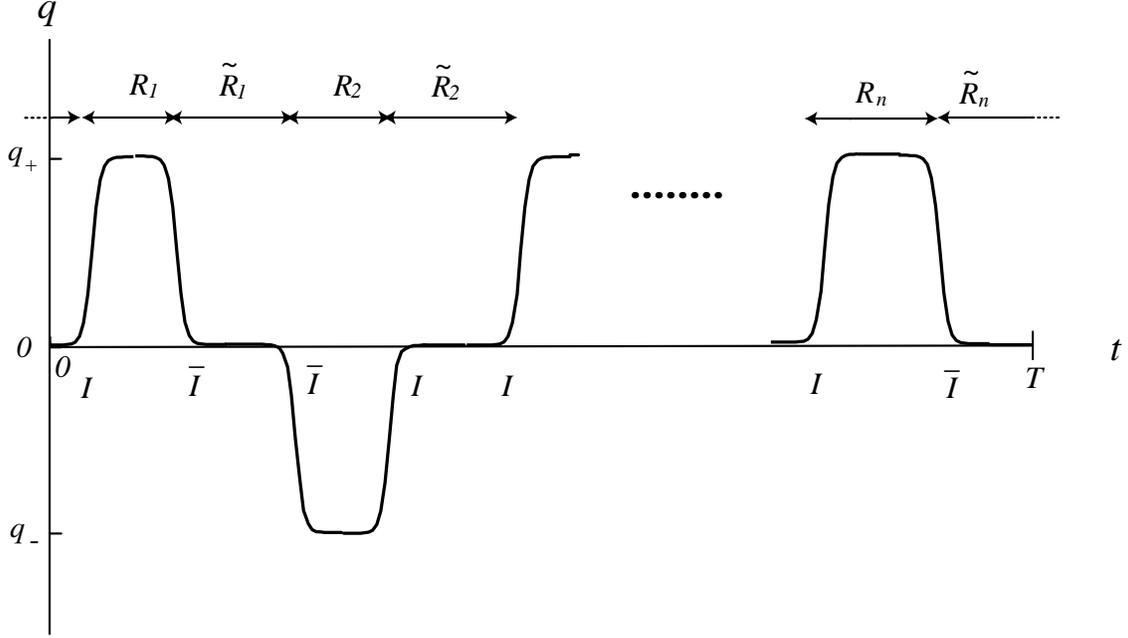}
\caption{The collective coordinates $R_{i}$ and $\tR_{i}$ for a
$2n$ valley-instantons configuration.}
\label{fig:mvccu}
\end{center}
\end{figure}

The evaluation of the partition function $Z=\tr e^{-HT}$ by
summing over multi-valley-instanton configurations can be
done in the same manner as those for the double-well and the
periodic potentials. One can easily see that in order to
incorporate with the periodic boundary condition in $T$, the number
of the valley-instantons in a period $T$ must be even. For a given
number $2n$ of the valley-instantons, however, there are still
several configurations. If we regard a configuration as $n$
valley-instanton pairs, we have four kinds of pair, $II$-,
$I\bar{I}$-, $\bar{I}I$-, and $\bar{I}\bar{I}$-valleys.
We denote the number of the $II$- and $\bar{I}\bar{I}$-valley
as $n_{II}$ and $n_{\bar{I}\bar{I}}$, respectively, and that
of the others as $n_{I\bar{I}}$. Contrary to the periodic
potential case, the particle must come back to the start point
after the period $T$ passes in this case. Therefore, we must
impose Eq.~(\ref{eqn:perbc1}) rather than Eq.~(\ref{eqn:perbc2}).
This condition results in $n_{II}=n_{\bar{I}\bar{I}}$.
As a consequence we have,
\begin{eqnarray}
2n_{II}+n_{I\bar{I}}=n.
\label{eqn:resnI}
\end{eqnarray}
This restriction shows that for a given $n$ there are $[n/2]+1$
variety of the $n_{II}$ value. For $n$ and $n_{II}$ fixed,
however, the configuration is not determined uniquely yet. There
remains a freedom of the permutation of the pairs. The number of
cases can be calculated if one notices that the configuration is
uniquely determined as far as the position of the $II$- and
$\bar{I}\bar{I}$-valleys among $n$ area is fixed. We denote
a set of the position as $\{i_{II}\}$. It is therefore clear
that for given $n$ and $n_{II}$ there are ${}_{n}C_{2n_{II}}$
configurations of the multiple valley-instantons.
Combining the results on well-separated valley-instanton pairs with
the above considerations, the well-separated multi-valley-instanton
action for given $n$, $n_{II}$ and $\{i_{II}\}$ is expressed as,
\begin{eqnarray}
S_{n, n_{II}}^{\{i_{II}\}}&=&2nS_{0}^{(I)}-\epsilon\sum_{i=1}^{n}R_{i}
 +\frac{\epsilon}{2}\sum_{i=1}^{n}\tR_{i}\nonumber\\
&& -\frac{1}{g^{2}}\sum_{i=1}^{n}e^{-2R_{i}}+\frac{2}{g^{2}}
 \sum_{i\in\{i_{II}\}}e^{-\tR_{i}}-\frac{2}{g^{2}}
 \sum_{i\notin\{i_{II}\}}e^{-\tR_{i}},
\label{eqn:actcu}
\end{eqnarray}
where $R_{i}$ is the distance between the $(2i-1)$-th and
$2i$-th (anti-)valley-instanton and $\tR_{i}$ the one between
the $2i$-th and the $(2i+1)$-th (anti-)valley-instanton$\pmod{n}$,
see Fig.~\ref{fig:mvccu}.

The sum of the contributions from the $2n$ valley-instantons
configuration can be written as,
\begin{eqnarray}
Z_{\rNP}=\sum_{n=1}^{\infty}\alpha^{2n}J_{n},\quad
 J_{n}=\sum_{n_{II}=0}^{[n/2]}\sum_{\{i_{II}\}}
 \cJ_{n, n_{II}}^{\{i_{II}\}},
\label{eqn:Znpcu}
\end{eqnarray}
where $\alpha^{2}$ denotes the contribution of the Jacobian and
the $R$-independent part of the determinant for one
valley-instanton-pair and is calculated as, in this case,
\begin{eqnarray}
\alpha^{2}=\frac{\sqrt{2}}{\pi g^{2}}e^{-1/2g^{2}}.
\label{eqn:defacu}
\end{eqnarray}
The term $\cJ_{n, n_{II}}^{\{i_{II}\}}$ is given by,
\begin{eqnarray}
\cJ_{n, n_{II}}^{\{i_{II}\}}&=&\frac{T}{n}\int_{0}^{\infty}\!
 \left(\prod_{i=1}^{n}dR_{i}\right)\left(\prod_{i=1}^{n}d\tR_{i}
 \right)\nonumber\\
&&\delta\left(\sum_{i=1}^{n}(R_{i}+\tR_{i})-T\right)\exp
 \left[ -( 1-\epsilon )\sum_{i=1}^{n}R_{i}
 -\left(\frac{1}{2}+\frac{\epsilon}{2}\right)\sum_{i=1}^{n}\tR_{i}
 \right.\nonumber\\
&&\left. +\frac{1}{g^{2}}\sum_{i=1}^{n}e^{-2R_{i}}-\frac{2}{g^{2}}
 \sum_{i\in\{i_{II}\}}e^{-\tR_{i}}+\frac{2}{g^{2}}
 \sum_{i\notin\{i_{II}\}}e^{-\tR_{i}}\right].
\label{eqn:dfJcu}
\end{eqnarray}
In the above expression, we notice that for $n$ and $n_{II}$
fixed, the contribution $\cJ_{n, n_{II}}^{\{i_{II}\}}$ does not
depend on the choice of the set $\{i_{II}\}$. This means the
following equality,
\begin{eqnarray}
\sum_{\{i_{II}\}}\cJ_{n, n_{II}}^{\{i_{II}\}}=\binom{n}{2n_{II}}
 \cJ_{n, n_{II}},
\label{eqn:newcJ}
\end{eqnarray}
where $\cJ_{n, n_{II}}$ is the contribution
$\cJ_{n, n_{II}}^{\{i_{II}\}}$ for a specific $\{i_{II}\}$
and is evaluated as,
\begin{eqnarray}
\cJ_{n, n_{II}}&=&\frac{T}{n}\int_{0}^{\infty}\!
 \left(\prod_{i=1}^{n}dR_{i}\right)\left(\prod_{i=1}^{n}d\tR_{i}
 \right)\nonumber\\
&&\delta\left(\sum_{i=1}^{n}(R_{i}+\tR_{i})-T\right)\exp
 \left[ -( 1-\epsilon )\sum_{i=1}^{n}R_{i}
 -\left(\frac{1}{2}+\frac{\epsilon}{2}\right)\sum_{i=1}^{n}\tR_{i}
 \right.\nonumber\\
&&\left. +\frac{1}{g^{2}}\sum_{i=1}^{n}e^{-2R_{i}}-\frac{2}{g^{2}}
 \sum_{i=1}^{2n_{II}}e^{-\tR_{i}}+\frac{2}{g^{2}}
 \sum_{i=2n_{II}+1}^{n}e^{-\tR_{i}}\right]\nonumber\\
&=&\frac{T}{2\pi in}\int_{-i\infty -\eta}^{i\infty -\eta}
 \! ds\, e^{-Ts}\, K_{-}(s)^{n}K_{+}^{(1)}(s)^{2n_{II}}
 K_{+}^{(2)}(s)^{n-2n_{II}}.
\label{eqn:dfnJcu}
\end{eqnarray}
In the last expression for $\cJ_{n, n_{II}}$, several $K$'s are
defined by,
\begin{subequations}
\label{eqns:defKs}
\begin{eqnarray}
K_{-}(s)&=&\int_{0}^{\infty}\! dR\exp\left[ ( s-1
 +\epsilon )R +\frac{1}{g^{2}}e^{-2R}\right]\nonumber\\
&\simeq&\frac{1}{2}\left( -\frac{1}{g^{2}}\right)^{\frac{s}{2}
 -\frac{1}{2}+\frac{\epsilon}{2}}\Gamma\left( -\frac{s}{2}
 +\frac{1}{2}-\frac{\epsilon}{2}\right),\\
K_{+}^{(1)}(s)&=&\int_{0}^{\infty}\! d\tR\exp\left[\left( s-\frac{1}{2}
 -\frac{\epsilon}{2}\right)\tR-\frac{2}{g^{2}}e^{-\tR}\right]\nonumber\\
&\simeq&\left(\frac{2}{g^{2}}\right)^{s-\frac{1}{2}
 -\frac{\epsilon}{2}}\Gamma\left( -s+\frac{1}{2}
 +\frac{\epsilon}{2}\right),\\
K_{+}^{(2)}(s)&=&\int_{0}^{\infty}\! d\tR\exp\left[\left( s-\frac{1}{2}
 -\frac{\epsilon}{2}\right)\tR+\frac{2}{g^{2}}e^{-\tR}\right]\nonumber\\
&\simeq&\left( -\frac{2}{g^{2}}\right)^{s-\frac{1}{2}
 -\frac{\epsilon}{2}}\Gamma\left( -s+\frac{1}{2}
 +\frac{\epsilon}{2}\right),
\end{eqnarray}
\end{subequations}
where the manipulation explained in subsection \ref{ssec:bogomo}
is again utilized for estimating each of the integrations.
Eventually, from Eqs.~(\ref{eqn:Znpcu}), (\ref{eqn:newcJ}) and
(\ref{eqn:dfnJcu}) we obtain,
\begin{eqnarray}
Z_{\rNP}&=&\sum_{n=1}^{\infty}\alpha^{2n}
 \sum_{n_{II}=0}^{[n/2]}{n\choose 2n_{II}}\cJ_{n, n_{II}}\nonumber\\
&=&-\frac{T}{4\pi i}\int_{-i\infty -\eta}^{i\infty -\eta}\! ds\,
 e^{-Ts}\ln\left( 1-\alpha^{2}K_{-}(s)K_{+}^{(+)}(s)\right)
 \left( 1-\alpha^{2}K_{-}(s)K_{+}^{(-)}(s)\right),
\label{eqn:Zcu1}
\end{eqnarray}
where,
\begin{eqnarray}
K_{+}^{(\pm)}(s)=K_{+}^{(2)}(s)\pm K_{+}^{(1)}(s).
\label{eqn:dfKpm}
\end{eqnarray}

\subsection{Non-perturbative Contributions}
\label{ssec:npccu}
From the results in Eqs.~(\ref{eqns:defKs})--(\ref{eqn:dfKpm}),
the non-perturbative contributions to
the spectra are determined by the following equation:
\begin{eqnarray}
\alpha^{2}\beta_{\pm}(E, \epsilon )\left(\frac{2}{g^{2}}
 \right)^{E-\frac{1}{2}-\frac{\epsilon}{2}}\Gamma\left(
 -E+\frac{1}{2}+\frac{\epsilon}{2}\right)
 \left(-\frac{1}{g^{2}}\right)^{\frac{E}{2}-\frac{1}{2}
 +\frac{\epsilon}{2}}\Gamma\left(-\frac{E}{2}+\frac{1}{2}
 -\frac{\epsilon}{2}\right)=1,
\label{eqn:spdcu}
\end{eqnarray}
where,
\begin{eqnarray}
\beta_{\pm}(E, \epsilon )=\frac{(-)^{E-\frac{1}{2}-\frac{\epsilon}{2}}
 \pm 1}{2}.
\label{eqn:defbcu}
\end{eqnarray}
We will solve the above equation by the series expansion in $\alpha$:
\begin{subequations}
\label{eqns:piacu}
\begin{eqnarray}
E_{n_{0}}&=&E_{n_{0}}^{(0)}+\alpha E_{n_{0}}^{(1)}
 +\alpha^{2} E_{n_{0}}^{(2)}+\cdots ,
 \quad E_{n_{0}}^{(0)}=n_{0}+\frac{1}{2}+\frac{\epsilon}{2},\\
E_{n_{\pm}}&=&E_{n_{\pm}}^{(0)}
 +\alpha E_{n_{\pm}}^{(1)}+\alpha^{2} E_{n_{\pm}}^{(2)}+\cdots ,
 \quad E_{n_{\pm}}^{(0)}=2n_{\pm}+1-\epsilon,
\end{eqnarray}
\end{subequations}
where $E_{n_{0}}$ stands for the spectra corresponding to, in the
limit $g\to 0$, the eigenfunctions of the center potential well and
$E_{n_{\pm}}$ for the ones corresponding to the parity eigenstates
obtained by the linear combinations of the eigenfunctions of
the each side potential well. For $\epsilon\ne\pm (2\cN +1)/3$
$(\cN =0,1,2,\dots )$,
the 1st order contributions vanish and the leading 2nd order
contributions are calculated as follows;
\begin{eqnarray}
E_{n_{0}}^{(2)}&=&-\frac{1}{n_{0}!}\left(
 \frac{2}{g^{2}}\right)^{n_{0}}\left(-\frac{1}{g^{2}}
 \right)^{\frac{n_{0}}{2}-\frac{1}{4}+\frac{3}{4}\epsilon}\Gamma
 \left(-\frac{n_{0}}{2}+\frac{1}{4}-\frac{3}{4}\epsilon\right),
\label{eqn:e2cu1}\\
E_{n_{\pm}}^{(2)}&=&-\left( (-)^{\frac{1-3\epsilon}{2}}\pm 1\right)
 \frac{1}{n_{\pm}!}\left(\frac{2}{g^{2}}
 \right)^{2n_{\pm}+\frac{1}{2}-\frac{3}{2}\epsilon}\left(
 \frac{1}{g^{2}}\right)^{n_{\pm}}\Gamma\left(-2n_{\pm}-\frac{1}{2}
 +\frac{3}{2}\epsilon\right).
\label{eqn:e2cu2}
\end{eqnarray}
In this case, degeneracies of the harmonic oscillator spectra
for the each potential well only occur between the both side wells.
The different non-perturbative contributions for $n_{\pm}$ in
Eq.~(\ref{eqn:e2cu2}) show the splitting of the degeneracies via the
quantum tunneling as in the case of symmetric double-well potentials.

\begin{figure}[ht]
\begin{center}
\includegraphics[width=.8\textwidth]{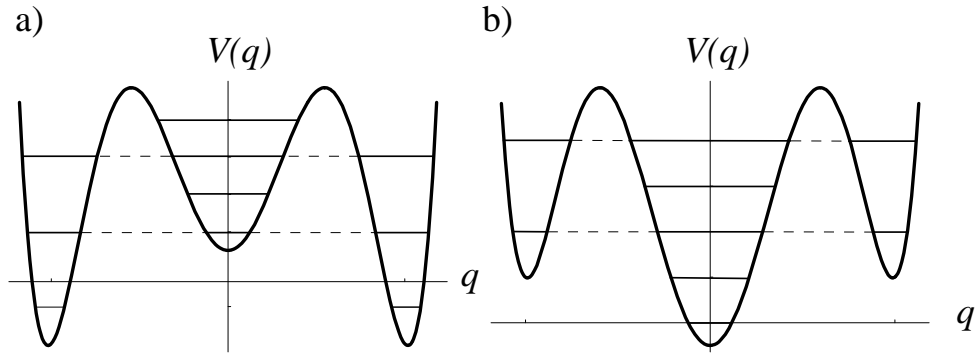}
\caption{Degeneracies of the harmonic spectra for (a)$\epsilon =5/3$
and (b)$\epsilon =-1$.}
\label{fig:dgcub1}
\end{center}
\end{figure}

When $\epsilon=(4\cN +1)/3$ $(\cN =0, 1, 2, \dots )$,
all the even-parity central harmonic spectra
$E_{2m_{0}}^{(0)}$ and the higher side harmonic spectra
$E_{n_{\pm}}^{(0)}$ degenerate for $n_{\pm}=m_{0}+\cN$,
see Fig.~\ref{fig:dgcub1}(a) for $\epsilon=5/3$ $(\cN =1)$.

It is interesting, however, that the interference due to
the quantum tunneling only occurs between the same (even-)parity
states. As a consequence, $E_{2m_{0}}$ and $E_{n_{+}}$
satisfying $n_{+}=m_{0}+\cN$ acquire order $\alpha^{1}$
contributions as follows:
\begin{eqnarray}
E_{n_{+}/2m_{0}}^{(1)}&=&\pm\sqrt{\frac{2}{n_{+}! (2m_{0})!}
 \left(\frac{2}{g^{2}}\right)^{2m_{0}}
 \left(\frac{1}{g^{2}}\right)^{n_{+}}},
\label{eqn:e1cu1}\\
E_{n_{+}/2m_{0}}^{(2)}&=&\frac{E_{n_{+}/2m_{0}}^{(1)\, 2}}{4}\left[
 \ln\left(-\frac{2}{g^{2}}\right)+\ln\left(\frac{2}{g^{2}}\right)
\right.\nonumber\\
&& \left. +\ln\left(-\frac{1}{g^{2}}\right)-\psi (n_{+}+1)
 -2\psi (2m_{0}+1)\right].
\label{eqn:e2cu3}
\end{eqnarray}
For the other spectra, say, $E_{n_{-}}$, $E_{2m_{0}+1}$ and the lower
$E_{n_{+}}$ with $n_{+}<\cN$, the contributions are the same as
Eqs.~(\ref{eqn:e2cu1}) and (\ref{eqn:e2cu2}). When $\epsilon =
-(4\cN -1)/3$ $(\cN =1, 2, 3, \dots )$, all the side harmonic spectra
$E_{n_{\pm}}^{(0)}$ and the higher even-parity central harmonic spectra
$E_{2m_{0}}^{(0)}$ degenerate for $m_{0}=n_{\pm}+\cN$, see
Fig.~\ref{fig:dgcub1}(b) for $\epsilon=-1$ $(\cN =1)$. In this case,
the interference also occurs only between the same (even-)parity states.
The contributions for $E_{2m_{0}}$ and $E_{n_{+}}$ satisfying
$m_{0}=n_{+}+\cN$ are given by the same as Eqs.~(\ref{eqn:e1cu1})
and (\ref{eqn:e2cu3}).
For the other spectra, say, $E_{n_{-}}$, $E_{2m_{0}+1}$ and the lower
$E_{2m_{0}}$ with $m_{0}<\cN$, the contributions are the same as
Eqs.~(\ref{eqn:e2cu1}) and (\ref{eqn:e2cu2}).

\begin{figure}[ht]
\begin{center}
\includegraphics[width=.8\textwidth]{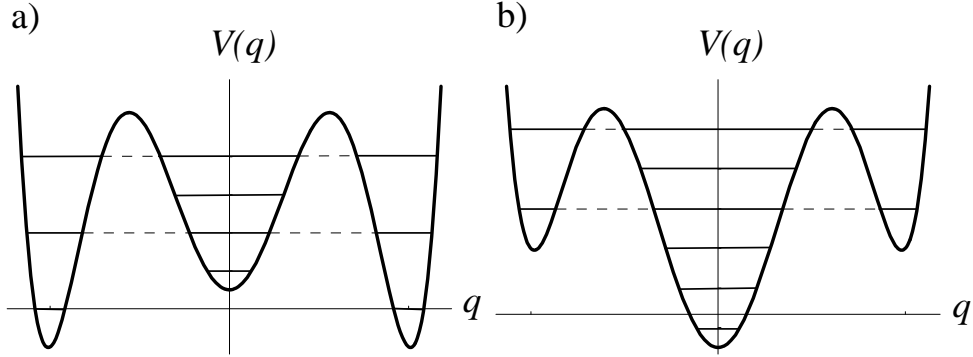}
\caption{Degeneracies of the harmonic spectra for (a)$\epsilon =1$
and (b)$\epsilon =-5/3$.}
\label{fig:dgcub2}
\end{center}
\end{figure}

When $\epsilon =(4\cN -1)/3$ $(\cN =1, 2, 3, \dots )$, all the
odd-parity central harmonic spectra $E_{2m_{0}+1}^{(0)}$ and the
higher side harmonic spectra $E_{n_{\pm}}^{(0)}$ degenerate
for $n_{\pm}=m_{0}+\cN$, see Fig.~\ref{fig:dgcub2}(a) for
$\epsilon=1$ $(\cN =1)$.

In this case, only the odd-parity states interfere and yield
order $\alpha^{1}$ contributions for $n_{-}=m_{0}+\cN$:
\begin{eqnarray}
E_{n_{-}/2m_{0}+1}^{(1)}&=&\pm\sqrt{\frac{2}{n_{-}! (2m_{0}+1)!}
 \left(\frac{2}{g^{2}}\right)^{2m_{0}+1}
 \left(\frac{1}{g^{2}}\right)^{n_{-}}},
\label{eqn:e1cu2}\\
E_{n_{-}/2m_{0}+1}^{(2)}&=&\frac{E_{n_{-}/2m_{0}+1}^{(1)\, 2}}{4}\left[
 \ln\left(-\frac{2}{g^{2}}\right)+\ln\left(\frac{2}{g^{2}}\right)
\right.\nonumber\\
&& \left. +\ln\left(-\frac{1}{g^{2}}\right)-\psi (n_{-}+1)-2
 \psi (2m_{0}+2)\right].
\label{eqn:e2cu4}
\end{eqnarray}
The contributions for the other spectra, say, $E_{n_{+}}$,
$E_{2m_{0}}$ and the lower $E_{n_{-}}$ with $n_{-}<\cN$, are
given by the same as Eqs.~(\ref{eqn:e2cu1}) and (\ref{eqn:e2cu2}).
When $\epsilon =-(4\cN +1)/3$ $(\cN =0, 1, 2, \dots )$,
all the side harmonic spectra $E_{n_{\pm}}^{(0)}$ and the higher
odd-parity central harmonic spectra $E_{2m_{0}+1}^{(0)}$
degenerate for $m_{0}=n_{\pm}+\cN$, see Fig.~\ref{fig:dgcub2}(b)
for $\epsilon=-5/3$ $(\cN =1)$.
Only the odd-parity states interfere in the same way
and the non-perturbative contributions
for $E_{2m_{0}+1}$ and $E_{n_{-}}$ satisfying $m_{0}=n_{-}+\cN$
are the same as Eqs.~(\ref{eqn:e1cu2}) and (\ref{eqn:e2cu4}).
Again, the expressions for the other spectra, say, $E_{n_{+}}$,
$E_{2m_{0}}$ and the lower $E_{2m_{0}+1}$ with $m_{0}<\cN$, are
given by Eqs.~(\ref{eqn:e2cu1}) and (\ref{eqn:e2cu2}).

From the whole result obtained here, we see that the non-perturbative
corrections vanish only when $\epsilon =(4\cN\pm 1)/3$
$(\cN =1,2,3,\dots )$. More precisely, when $\epsilon =(4\cN +1)/3$,
Eq.~(\ref{eqn:e2cu2}) is applied for the even-parity
states labeled by the quantum number $n_{-}$ and results in
$E_{n_{-}}^{(2)}=0$ for all $n_{-}<\cN$. Similarly, when
$\epsilon =(4\cN -1)/3$, Eq.~(\ref{eqn:e2cu2}) is applied for the
odd-parity states labeled by the quantum number $n_{+}$ and
results in $E_{n_{+}}^{(2)}=0$ for all $n_{+}<\cN$. It should be noted
that in the case of $\epsilon=-(4\cN\pm 1)/3$ $(\cN =1,2,3,\dots )$
the non-perturbative corrections do remain although the models are
$\cN$-fold supersymmetric, reflecting the fact that they are only
quasi-perturbatively solvable but are not quasi-exactly solvable.
These results are just what we have expected from the general
properties of $\cN$-fold supersymmetry.

\subsection{Large Order Behavior of the Perturbation Series}
\label{ssec:lobcu}
The large order behavior of the perturbation series in $g^{2}$
for the spectra can be estimated by the same way as in the case
of the double-well and periodic potentials. From the non-perturbative
contributions Eqs.~(\ref{eqn:e2cu1})--(\ref{eqn:e2cu4}), we can
easily see that the imaginary parts of them are continuous, at
least up to the order $\alpha^{2}$, in $\epsilon$ and yield,
\begin{subequations}
\label{eqns:imecus}
\begin{eqnarray}
\rIm E_{n_{0}}&\sim&-\alpha^{2}\frac{\pi}
 {n_{0}!\, \Gamma \left(\frac{n_{0}}{2}+\frac{3}{4}+\frac{3}{4}
 \epsilon \right)}\left(\frac{2}{g^{2}}\right)^{n_{0}}
 \left(\frac{1}{g^{2}}\right)^{\frac{n_{0}}{2}-\frac{1}{4}
 +\frac{3}{4}\epsilon},\\
\rIm E_{n_{\pm}}&\sim&-\alpha^{2}\frac{\pi}
 {n_{\pm}!\, \Gamma \left( 2n_{\pm}+\frac{3}{2}-\frac{3}{2}\epsilon
 \right)}\left(\frac{2}{g^{2}}\right)^{2n_{\pm}+\frac{1}{2}
 -\frac{3}{2}\epsilon}\left(\frac{1}{g^{2}}\right)^{n_{\pm}},
\end{eqnarray}
\end{subequations}
which are valid for arbitrary $\epsilon$. Then, if we expand
the spectra in power of $g^{2}$ such that,
\begin{eqnarray}
E_{n_{0}/n_{\pm}}=E_{n_{0}/n_{\pm}}^{(0)}+\sum_{r=1}^{\infty}
 a_{n_{0}/n_{\pm}}^{(r)}g^{2r},
\label{eqn:dfptcu}
\end{eqnarray}
the large order behavior of the coefficients $a^{(r)}$ for
sufficiently large $r$ are calculated as, using Eqs.~(\ref{eqn:disrel})
and (\ref{eqns:imecus}),
\begin{subequations}
\label{eqns:lobcus}
\begin{eqnarray}
a_{n_{0}}^{(r)}&\sim&A_{n_{0}}(\epsilon)\, 2^{r}\,\Gamma\left( r
 +\frac{3}{2}n_{0}+\frac{3}{4}+\frac{3}{4}\epsilon\right),\\
a_{n_{\pm}}^{(r)}&\sim&A_{n_{\pm}}(\epsilon)\, 2^{r}\,\Gamma\left(
 r+3n_{\pm}+\frac{3}{2}-\frac{3}{2}\epsilon\right),
\end{eqnarray}
\end{subequations}
where,
\begin{subequations}
\label{eqns:lobpcus}
\begin{eqnarray}
A_{n_{0}}(\epsilon)&=&\frac{\sqrt{2}}{\pi}\frac{2^{\frac{5}{2}n_{0}
 +\frac{3}{4}+\frac{3}{4}\epsilon}}{n_{0}!\,\Gamma\left(
 \frac{n_{0}}{2}+\frac{3}{4}+\frac{3}{4}\epsilon\right)},\\
A_{n_{\pm}}(\epsilon)&=&\frac{\sqrt{2}}{\pi}\frac{2^{5n_{\pm}
 +2-3\epsilon}}{n_{\pm}!\,\Gamma\left(
 2n_{\pm}+\frac{3}{2}-\frac{3}{2}\epsilon\right)}.
\end{eqnarray}
\end{subequations}
Equations (\ref{eqns:lobcus}) show that the perturbative
coefficients diverge factorially unless the prefactor $A(\epsilon )$'s
vanish. From Eq.~(\ref{eqns:lobpcus}), we can find the disappearance
of the leading divergence takes place only when $\epsilon =\pm (2n+1)/3$
$(n=1, 2, 3, \dots )$. More precisely, we obtain the following results:
\renewcommand{\labelenumi}{(\arabic{enumi})}
\begin{enumerate}
\item $\epsilon =(4\cN\pm 1)/3\, (\cN =1, 2, 3, \dots )$
 \[
  A_{n_{\pm}}(\epsilon )=0 \quad\text{for}\quad n_{\pm}<\cN .
 \]
\item $\epsilon =-(4\cN +1)/3\, (\cN =1, 2, 3, \dots )$
 \[
  A_{2m_{0}+1}(\epsilon )=0 \quad\text{for}\quad m_{0}<\cN .
 \]
\item $\epsilon =-(4\cN -1)/3\, (\cN =1, 2, 3, \dots )$
 \[
  A_{2m_{0}}(\epsilon )=0 \quad\text{for}\quad m_{0}<\cN .
 \]
\end{enumerate}
Comparing these results with Eq.~(\ref{eqn:epvnfc}), we see
that the above cases completely coincide with the case where
the system possesses Type A $\cN$-fold supersymmetry. Again,
the results of the valley method analyses are consistent with
the non-renormalization theorem.

\section{Summary}
\label{sec:concl}
In this article, we have made non-perturbative analyses on the models
which can be $\cN$-fold supersymmetric at specific values of the
parameter. Combining the results obtained in this article with the ones
in Ref.~\cite{AKOSW2}, we get the following:
\begin{enumerate}
\item For all the potentials investigated (double-well, triple-well,
 periodic), the leading divergence of the perturbation series disappears
 when and only when they are $\cN$-fold supersymmetric. The
 non-renormalization theorem ensures that $\cN$-fold supersymmetry
 is sufficient for the disappearance of the divergence. The results
 indicate that it may also be necessary.
\item The non-perturbative corrections to the spectra for certain
 states vanish when and only when the models are quasi-exactly
 solvable (triple-well, periodic).
\item For the quasi-perturbatively solvable potentials (double-well,
 triple-well), the non-perturbative corrections remain although
 they are $\cN$-fold supersymmetric.
\end{enumerate}
As was mentioned in Ref.~\cite{Turb} the quasi-solvable models
constructed by $sl(2)$ generators do not always have normalizable
solvable states. Although the conditions on the normalizability
of the models were fully investigated in Ref.~\cite{GLKO},
it remains unclear what is the role of the partial algebraization
of the models without normalizable solvable states. The results
listed above provide an answer to this problem. Even though
the solvable wave functions are not normalizable, they can be
normalizable and thus make sense in the perturbation theory. 
In this case, the spectra corresponding to the solvable states
also make sense in the perturbation theory. As was shown in
Ref.~\cite{AST2}, the perturbation series for them are convergent
since they are the solutions of a finite order algebraic equation.
However, the fact that the solvable states and the corresponding
spectra make sense only in the perturbation theory inevitably
means the existence of the non-perturbative effects, which is
in contrast to the case of the quasi-exactly solvable models.
That is why we have called the case quasi-perturbatively solvable.

Finally, we would like to mention about applicability of
the dilute-gas approximation. As has been mentioned previously,
the dilute-gas approximation cannot give proper results, that is,
consistent results with $\cN$-fold supersymmetry, for both the
potentials Eqs.~(\ref{eqn:exppo}) and (\ref{eqn:cubpo}).
Therefore, it seems that the success of the dilute-gas approximation
for the symmetric double-well potential is rather exceptional and
applicability of it is quite limited.

\begin{acknowledgments}%REVTEX4
%\begin{ack}%ELSART

We would like to thank H.~Aoyama for a useful discussion.
T.~Tanaka's work was supported in part by a JSPS research fellowship.

\end{acknowledgments}%REVTEX4
%\end{ack}%ELSART

%\appendix

\end{document}